\documentclass{article}

\usepackage{float}
\usepackage{lipsum}
\usepackage{amsfonts}
\usepackage{graphicx}
\usepackage{epstopdf}
\usepackage{algorithm}
\usepackage{algorithmic}

\title{A Quasi-Polynomial Approximation for the Restricted Assignment Problem\thanks{
This article is an extended joint version of conference articles \cite{DBLP:conf/soda/JansenR17, DBLP:conf/ipco/JansenR17}.
This research was supported by German Research Foundation (DFG) project JA 612/15-1.}}

\author{Klaus Jansen \quad Lars Rohwedder \medskip\\
  Department of Computer Science, University of Kiel, Germany \\
  \{kj, lro\}@informatik.uni-kiel.de}

\usepackage{amsopn}
\usepackage{amsmath}
\usepackage{amssymb}
\usepackage{amsthm}
\usepackage{tikz}

\newcommand\jobs{\mathcal J}
\newcommand\machines{\mathcal M}
\newcommand\OPT{\mathrm{OPT}}

\newtheorem{theorem}{Theorem}
\newtheorem{lemma}[theorem]{Lemma}
\newtheorem{definition}[theorem]{Definition}
\newtheorem{claim}[theorem]{Claim}

\begin{document}

\maketitle

\begin{abstract}
The Restricted Assignment Problem is a prominent special case of Scheduling on Parallel
Unrelated Machines.
For the strongest known linear programming relaxation, the configuration LP,
we improve the non-constructive bound
on its integrality gap from $1.9142$ to $1.8334$ and significantly simplify the proof.
Then we give a constructive variant, yielding a
$1.8334$-approximation in quasi-polynomial time.
This is the first quasi-polynomial algorithm for this problem
improving on the long-standing approximation rate of $2$.
\end{abstract}

\section{Introduction}
We consider a special case of the problem \textsc{Scheduling on Unrelated Parallel Machines},
where the goal is to compute an allocation $\sigma: \mathcal J\rightarrow \mathcal M$
of the jobs $\mathcal J$ to the machines $\mathcal M$.
On machine $i$ the job $j$ has a processing time (size) $p_{ij}$.
We want to minimize the makespan, which is the maximum load
$\max_{i\in\mathcal M}\sum_{j\in\sigma^{-1}(i)} p_{ij}$.
The classical 2-approximation by Lenstra et al.~\cite{DBLP:journals/mp/LenstraST90} 
is still the algorithm of choice for this problem. They also show that
no approximation ratio better than $3/2$ can be found in polynomial time unless $\mathrm{P} = \mathrm{NP}$.
Closing this gap appears in several lists of important open questions:
Schuurman and Woeginger~\cite{schuurman1999polynomial} include it in their influential
survey on open questions scheduling and 
Shmoys and Williamson in their book on approximation algorithms~\cite{DBLP:books/daglib/0030297}.

While the general problem remains unclear, there has been
progress on a special case called \textsc{Restricted Assignment}.
Here each job $j$ has a processing time $p_j$, which is independent from the
machines, and a set of feasible machines $\Gamma(j)$. This means $j$
can only be assigned to one of the machines in $\Gamma(j)$.
Note that this is equivalent to the previous problem when $p_{ij}\in\{p_j,\infty\}$.
The lower bound of $3/2$ holds also in the restricted case and
even if given quasi-polynomial running time no better approximation ratio can be obtained,
unless $\mathrm{DTIME}(2^{\mathrm{polylog}(n)}) = \mathrm{NP}$,
which would contradict popular conjectures such as the Exponential Time Hypothesis.

In a seminal work, Svensson~\cite{DBLP:journals/siamcomp/Svensson12} 
proved that the configuration LP, a natural
linear programming relaxation,
has an integrality gap of at most $33/17$.
By approximating the optimum of the configuration LP this yields an $(33/17 + \epsilon)$-estimation algorithm.
However, this proof is non-constructive and no polynomial
algorithm is known that can produce a solution of this quality.

For instances with only two processing times additional progress has been made.
Chakrabarty et al. gave a polynomial $(2-\delta)$-approximation
for a very small $\delta$~\cite{DBLP:conf/soda/ChakrabartyKL15}.
Annamalai improved this with a
$(17/9 + \epsilon)$-approximation for every $\epsilon > 0$~\cite{DBLP:journals/corr/Annamalai16}.
For this special case it was also shown that the integrality gap is at
most $5/3$~\cite{DBLP:conf/swat/JansenLM16}.

In \cite{DBLP:journals/siamcomp/Svensson12} and \cite{DBLP:conf/swat/JansenLM16}
the critical idea is to design a local search algorithm,
which is then shown to produce good solutions. However, the algorithm has a potentially high
running time; hence it could only be used to prove the existence of a good solution.
For the closely related \textsc{Santa Claus} problem, in which the minimum is maximized
instead of the maximum being minimized, similar algorithms were developed~\cite{DBLP:journals/talg/AsadpourFS12}.
There, a quasi-polynomial variant by Pol{\'{a}}cek et al.~\cite{DBLP:journals/talg/PolacekS16}
and a polynomial variant by Annamalai et al.~\cite{DBLP:journals/talg/AnnamalaiKS17} were later discovered.

In this article, we start by giving a simple quasi-polynomial time $(2+\epsilon)$-approximation
in order to introduce some ideas.
We then present a simpler variant of Svensson's non-constructive algorithm, which in
addition achieves a better approximation ratio; thereby we improve the bound on the
integrality gap of the configuration LP.
Finally, we combine both approaches in a
very sophisticated $(11/6+\epsilon)$-approximation algorithm that terminates in quasi-polynomial time.
This leads to the first better-than-2 approximation algorithm
for \textsc{Restricted Assignment}, which does not need exponential running time.
The algorithm is purely combinatorial and uses the configuration-LP only in the analysis.
The last algorithm implies the results of the previous two.
Nevertheless, they are significantly less complex and already
demonstrate many of the key techniques used in the last algorithm.
\paragraph*{Comparison to related algorithms}
In Svensson's algorithm~\cite{DBLP:journals/siamcomp/Svensson12} and ours,
jobs are moved until the desired allocation is found.
The presentations of the algorithms differ significantly, but on a high level Svensson's
algorithm is closely related to the second (exponential time) algorithm we give.
Considering simplified instances with only jobs of two sizes, both algorithms and
their analysis are essentially the same.
Our exponential time algorithm is a cleaner adaption to general instances,
which is much less technical and has an better approximation ratio. 

The approach for obtaining a quasi-polynomial
running time resembles that of~\cite{DBLP:journals/talg/PolacekS16}, where it was done
for the restricted \textsc{Santa Claus} problem.
In the \textsc{Restricted Assignment} problem, however,
this turns out to be significantly more challenging.
The basic idea in both algorithms is to reduce the search depth to a logarithmic
value. A fundamental structure in both cases are chains of big jobs or resources.
Big jobs have a size greater than $1/2$ times the optimal makespan.
Informally, a chain of big jobs is a sequence $j_1,i_1,j_2,i_2,j_3,i_3\dotsc$, where $j_1$ is
a big job allowed to be placed on $i_1$; $j_2$ is a big job currently assigned to $i_1$, which is
also allowed on $i_2$; $j_3$ is a big job currently assigned to $i_2$, but also allowed on $i_3$; etc.
A similar situation can arise in the restricted \textsc{Santa Claus} problem, except
that big resources (the counter-part to jobs) have a size at least the desired solution value.
It turns out that these chains play a critical role and in the \textsc{Santa Claus} problem they
have a very simple structure. This is because on a player (the counter-part to a machine)
which has one big resource we do not place any other resources.
In the \textsc{Restricted Assignment} case it is necessary to place also other jobs on 
machines that have big jobs. This means that the simple operation of moving every job of a chain
to the next machine works well in the \textsc{Santa Claus} case, but in the
\textsc{Restricted Assignment} case this can result in bad machines (machines that
have too much load), if we are not careful.
Perhaps this is also a reason why at this time there is no polynomial time
better-than-$2$ approximation algorithm known for \textsc{Restricted Assignment} and
the only progress in this direction is in the case where we have only one big job size.
Note that in this case, chains are again simple.
\paragraph*{Notation}
For a set of jobs $A\subseteq\jobs$, we write $p(A)$ in place
of $\sum_{j\in A} p_j$. For other variables indexed by jobs,
we may do the same.
An allocation is a function $\sigma: \jobs\rightarrow\machines$,
where $\sigma(j)\in\Gamma(j)$ for all $j\in\jobs$.
We write $\sigma^{-1}(i)$ for the set of 
all jobs $j$ which have $\sigma(j) = i$.
\paragraph*{Configuration LP}
The configuration LP has an exponential size, but can be
solved approximately in polynomial time with a rate of $(1 + \epsilon)$
for every $\epsilon > 0$~\cite{DBLP:conf/stoc/BansalS06}.
For every machine $i$ and every $\tau \ge 0$ let
\begin{equation*}
  \mathcal C(i, \tau) = \{ S \subseteq \jobs : p(S) \le \tau \text{ and for all $j\in S$}, i\in\Gamma(j)\} .
\end{equation*}
These are the configurations for machine $i$ and makespan $\tau$. They are a set of
jobs that have volume at most $\tau$ and can run on machine $i$.
The optimum $\OPT^*$ of the configuration LP is the lowest $\tau$ such that the following linear
program is feasible.
\begingroup
\floatname{algorithm}{Linear Program}
\begin{algorithm}
\caption{Primal of the configuration LP}
\begin{align*}
  \sum_{C\in\mathcal C(i, \tau)} x_{i, C} &\le 1 & \forall i\in\machines \\
  \sum_{i\in\machines}\sum_{C\in\mathcal C(i,\tau) : j\in C} x_{i, C} &\ge 1 & \forall j\in\jobs \\
  x_{i, C} &\ge 0
\end{align*}
\end{algorithm}
This linear program assigns at least one configuration to every machine and makes sure that
every job is assigned at least once. We will also construct the dual after
adding the objective $\max \ (0,\dotsc,0) \cdot x$ to the configuration LP:
\begin{algorithm}
\caption{Dual of the configuration LP}

\begin{align*}
  \min \sum_{i\in\machines} y_i &- \sum_{j\in\jobs} z_j \\
  \sum_{j\in C} z_j &\le y_i &\forall i\in\machines, C\in\mathcal C(i, \tau) \\
  y_i, z_j &\ge 0
\end{align*}
\end{algorithm}
\endgroup
Recall, the value $\tau$ is a constant in the LP and, if the configuration LP is
infeasible with $\tau$, this means $\OPT^* > \tau$.
Furthermore, we can derive the following condition from duality.
\begin{lemma}\label{lemma:condition-ra}
  Let $y\in \mathbb R_{\ge 0}^\machines$ and $z\in \mathbb R_{\ge 0}^\jobs$ such
  that $\sum_{i\in\machines} y_i < \sum_{j\in\jobs} z_j$ and
  for every $i\in\machines$ and $C\in\mathcal C(i, \tau)$ it holds
  that $\sum_{j\in C} z_j \le y_i$, then
  $\OPT^* > \tau$.
\end{lemma}
It is easy to see that if such a solution $y, z$ exists, then every component can be scaled
by a constant to obtain a feasible solution lower than any given value. Hence, the 
dual must be unbounded and therefore the primal must be infeasible.
\section{Simple algorithm}
In this section we present a quasi-polynomial time $(2+\epsilon)$-approximation algorithm.
It should be noted that there do exist clean and simple polynomial time $2$-approximation algorithms
for this problem and even for the general problem of scheduling on unrelated machines.
The purpose of this section is merely to introduce some concepts
and how they can be used to get a quasi-polynomial running time.
We use a dual approximation framework where we perform a binary search over variable
$\tau\in [p_{\max},\ n\cdot p_{\max}]$, where $p_{\max} = \max_{j\in\jobs} p_j$. In each iteration we either prove that $\OPT^* > \tau$ or find an allocation
with makespan at most $(1+\epsilon)\tau$. We stop the binary search once upper and lower bound
differ by less than a factor of $(1+\epsilon)$. This gives a $(1+\epsilon)^2$-approximation
in $\log_{1+\epsilon}(n) = O(1/\epsilon\cdot\log(n))$ iterations of the binary search.
By scaling down $\epsilon$ we can get a $(1+\epsilon)$-approximation in the same asymptotic time.
The algorithmic idea for the inner method is basically a breadth-first search:

Let $\sigma: \jobs\rightarrow\machines$ be an arbitrary allocation.
We use layers $L_0,\dotsc,L_\ell$, which are disjoint sets of machines. We write $L_{\le k}$
for the union over all machines in $L_0,\dotsc,L_k$. 
Further, let $\tilde\jobs(L_{\le k}, \sigma)$ denote the set of jobs $j$ with $\sigma(j)\in L_{\le k}$.
We call a machine $i$ \emph{good}, if $p(\sigma^{-1}(i)) \le 2 + \epsilon$, and \emph{bad}, otherwise.
The algorithm (see Alg.~\ref{alg:simple}) initializes $L_0$ as the bad machines. Then
the subsequent layers $L_{k+1}$ are created as the union over all $\Gamma(j)\setminus L_{\le k}$ where
$j\in\tilde\jobs(L_{\le k}, \sigma)$. If there is a machine with a load at most $(1+\epsilon)\tau$
in some layer, we move a job from a lower layer to this layer and then start from the beginning.

\begin{algorithm}
\caption{Quasi-polynomial $(2+\epsilon)$-approximation algorithm}
\label{alg:simple}

\begin{algorithmic}
\STATE{let $\sigma$ be an arbitrary allocation}
\STATE{$\ell\gets 0$}
\STATE{let $L_0$ be the set of bad machines}
\WHILE{$L_0\neq\emptyset$}
\STATE{let $L_{\ell+1}$ be the union over all $\Gamma(j)\setminus L_{\le \ell}$ for $j\in\tilde\jobs(L_{\le i}, \sigma)$}
\IF{there is an $i\in L_{\ell+1}$ with $p(\sigma^{-1}(i)) \le (1+\epsilon)\tau$}
  \STATE{find a job $j$ with $\sigma(j)\in L_{\le \ell}$ and $i\in\Gamma(j)$}
  \STATE{$\sigma(j)\leftarrow i$}
  \STATE{delete $L_0,\dotsc,L_{\ell+1}$}
  \STATE{$\ell\gets 0$}
  \STATE{let $L_0$ be the new set of bad machines}
\ELSE
  \STATE{$\ell\gets\ell+1$}
\IF{$\ell \ge \log_{1+\epsilon}(|\machines|)$}
  \RETURN "err"
\ENDIF
\ENDIF
\ENDWHILE
\end{algorithmic}
\end{algorithm}

\paragraph*{Running time} We consider two consecutive iterations
right before some job is moved and the layers are deleted. Let $\sigma$ and $\sigma'$ be
the allocations at the earlier and at the later iteration. Likewise, let $L_{\le\ell}$ and
$L'_{\le\ell'}$ be the layers.
We show that the vector
\begin{equation*}
  (b', |\tilde\jobs(L'_{\le 0}, \sigma)|,\dotsc,|\tilde\jobs(L'_{\le \ell'}, \sigma)|,-1)
\end{equation*}
is lexicographically smaller than
\begin{equation*}
  (b, |\tilde\jobs(L_{\le 0}, \sigma)|,\dotsc,|\tilde\jobs(L_{\le \ell}, \sigma)|,-1) ,
\end{equation*}
where $b'$ and $b$ are the number of bad machines for $\sigma'$ and $\sigma$.
Since the number of layers is at most
$\log_{1+\epsilon}(|\machines|) = O(1/\epsilon\cdot\log(|\machines|))$ and each component is bounded
by $|\jobs|$, the overall running time is $|\jobs|^{O(1/\epsilon\cdot\log(|\machines|))}$.
For the lexicographic decrease, notice that the algorithm moves a job $j$ to a machine $i$
only when $p(\sigma^{-1}(i))\le (1+\epsilon)\tau$. Hence, after adding $j$, the load on $i$
is $p(\sigma^{-1}(i)) + p_j \le (1+\epsilon)\tau + \tau \le (2+\epsilon)\tau$.
Thus, the algorithm never turns a good machine into a bad machine and $b'\le b$.
If $b' < b$ we are done. Otherwise, $b'=b$ and
since no jobs were moved to or from $L_{\le\ell-1}$,
$L'_{\le k} = L_{\le k}$ and $\tilde\jobs(L'_{\le k}, \sigma') = \tilde\jobs(L_{\le k}, \sigma)$ for all
$k\le \min\{\ell-1, \ell'\}$.
If $\ell'\le\ell-1$, then the $\ell'+1$-th component will be $-1$ and therefore smaller than
$|\tilde\jobs(L_{\le\ell'+1},\sigma)|$. Otherwise, the $\ell$-th component will be smaller because
we moved one job away from $L_{\le\ell}$.
\paragraph*{Correctness} We have to verify that if the algorithm returns "err", then $\OPT^* > \tau$.
We will do so using Lemma~\ref{lemma:condition-ra}.
Assume that $\ell\ge \log_{1+\epsilon}(|\machines|)$ and let $\sigma$ and $L_{\le\ell}$ be the 
current allocation and layer structure.
For every $i\in L_k$ define
\begin{equation*}
  y_i = (1+\epsilon)^{1-k} .
\end{equation*}
Furthermore, define $y_i = (1+\epsilon)^{-\log_{1+\epsilon}(|\machines|)}$, if $i\notin L_{\le\ell}$.
For jobs $j\in\jobs$ set
\begin{equation*}
  z_j = (1+\epsilon)^{-k} \cdot p_j/\tau,
\end{equation*}
where $k$ is minimal with $j\in\tilde\jobs(L_{\le k}, \sigma)$ and $z_j = 0$ if there is no
such $k$.
We need to show that $\sum_{j\in\jobs} z_j > \sum_{i\in\machines} y_i$ and for all
$i\in\machines$ and $C\in\mathcal C(i, \tau)$ it holds that $z(C)\le y_i$.
For the former, we argue
\begin{multline*}
  \sum_{j\in\jobs} z_j = \sum_{i\in\machines} z(\sigma^{-1}(i))
  = \sum_{k=0}^{\ell} \sum_{i\in L_k} (1+\epsilon)^{-k} \frac{p(\sigma^{-1}(i))}{\tau} \\
  > (2+\epsilon) |L_0| + \sum_{k=1}^{\ell} (1+\epsilon)^{-k} (1+\epsilon) |L_k|
  \ge 1 + \sum_{k=0}^{\ell} (1+\epsilon)^{1-k} |L_k| \\
  = |\machines| \cdot (1+\epsilon)^{-\log_{1+\epsilon}(|\machines|)} + \sum_{k=0}^{\ell} (1+\epsilon)^{1-k} |L_k|
  \ge \sum_{i\in\machines} y_i .
\end{multline*}
For the latter condition, first consider a machine $i\in L_k$ and $C\in\mathcal C(i,\tau)$. There can be no job $j\in C$
with $j\in\tilde\jobs(L_{\le k-2}, \sigma)$, since otherwise $i$ would be in an earlier layer.
Therefore, $z_j \le (1+\epsilon)^{-(k-1)} p_j$ for all $j\in C$.
Thus,
\begin{equation*}
  z(C) \le (1+\epsilon)^{1-k} \frac{p(C)}{\tau} \le (1+\epsilon)^{1-k} = y_i .
\end{equation*}
Now let $i\in\machines\setminus L_{\le\ell}$ and let $C\in\mathcal C(i,\tau)$.
No job in $C$ can be in $\tilde\jobs(L_{\le \ell-1}, \sigma)$.
This means, $z_j \le (1+\epsilon)^{-\ell} p_j$ for all $j\in C$ and
\begin{equation*}
  z(C) \le (1+\epsilon)^{-\ell} \frac{p(C)}{\tau} \le (1+\epsilon)^{-\log_{1+\epsilon}(|\machines|)} = y_i .
\end{equation*}
Using the lemma this implies that $\OPT^* > \tau$.

\section{Non-constructive integrality gap bound}
In this section, we give an approximation algorithm with ratio $11/6$. The algorithm
is similar to the previous one, but it adds only single moves, instead of a whole layers of reachable
machines. This leads to an exponential running time bound. Hence, the algorithm in this section
only gives a non-constructive bound on the integrality gap of the configuration LP.

\subsection{Algorithm}
Given an allocation $\sigma : \jobs \rightarrow \machines$, we call a machine $i$
\emph{bad}, if $p(\sigma^{-1}(i)) > 11/6 \cdot \tau$.
A machine is \emph{good}, if it is not bad.
We define \emph{big} jobs to be those $j\in\jobs$ that have
$p_j > 1/2 \cdot \tau$ and small jobs all others.

As the previous one, this algorithm starts with an arbitrary allocation and moves jobs until
all machines are good, or it can prove that the configuration LP is
infeasible w.r.t. $\tau$. During this process, a machine that is already good
will never be made bad.

The central data structure of the algorithm is an ordered list of pending moves $P = (P_1,P_2,\dotsc,P_\ell)$.
Here, every component $P_k = (j, i)$, $j\in\jobs$ and $i\in\Gamma(j)$, stands for
a move the algorithm wants to perform.
It will not perform the move, if this would create a bad machine,
i.e., $p(\sigma^{-1}(i)) + p_j > 11/6 \cdot \tau$.
If it does not create a bad machine, we say that the move $(j, i)$ is valid.
For every $0\le k \le \ell$ define $L_{\le k} := (L_1,\dotsc,L_k)$, the first $k$ elements of $L$ (with $L_{\le 0}$ being the empty list).

Depending on the current allocation $\sigma$ and list of pending moves $P_{\le \ell}$,
we define a binary relation $R(P_{\le \ell}, \sigma)\subseteq \jobs\times\machines$.
For a pair $(j, i)\in R(P_{\le \ell}, \sigma)$ we say
machine $i$ \emph{repels} $j$ w.r.t. $P_{\le\ell}$. This does not
mention $\sigma$ and therefore slightly abuses notation, but during
the lifetime of $P_{\le\ell}$ the allocation $\sigma$ does
not change and is always clear from the context.
 
The definition of repelled jobs is given later.
The algorithm will only add a new move $(j, i)$ to the current list $P$, if 
$j$ is repelled by its current machine and
not repelled by the target $i$ w.r.t. $P$ (see Alg.~\ref{alg:ra}).

In the algorithm we use a lexicographic order $(p_j, j, i)$ of the moves $(j, i)$.
Here, we assume that there is an arbitrary order on jobs and machines, which is consistent throughout
the iterations.
\begin{algorithm}
\caption{Algorithm for \textsc{Restricted Assignment}}
\label{alg:ra}
\begin{algorithmic}
\STATE{let $\sigma$ be an arbitrary allocation}
\STATE{$\ell \gets 0$}
\WHILE{there is a bad machine}
  \STATE{choose a move $(j, i)\notin P_{\le\ell}$, $j\in\jobs$ and $i\in\Gamma(j)$, where
         $j$ is repelled by $\sigma(j)$ and not repelled by $i$ w.r.t. $P_{\le\ell}$
         and $(p_j, j, i)$ is lexicographically minimal among all candidates}
  \STATE{$P_{\ell+1} \leftarrow (j, i)$}
  \STATE{$\ell = \ell + 1$}
  \IF{$p(\sigma^{-1}(i) + p_j \le 11/6\cdot\tau$}
    \STATE{$\sigma(j) \leftarrow i$}
    \STATE{delete $P_{1},\dotsc,P_{\ell}$}
    \STATE{$\ell \leftarrow 0$}
  \ENDIF
\ENDWHILE
\end{algorithmic}
\end{algorithm}
\paragraph*{Repelled jobs}
We define the repelled jobs of each machine inductively w.r.t. $P_{\le k}$, $k=0,1,\dotsc,\ell$.
\begin{description}
  \item[(initialization)] If $k=0$, let every bad machine $i$ repel every job $j$ w.r.t. $P_{\le k}$.
  \item[(monotonicity)] If $i$ repels $j$ w.r.t. $L_{\le k}$,
    then let $i$ repel $j$ also w.r.t. $P_{\le k+1}$.
\end{description}
The remaining rules regard $k > 0$ and we let $(j_k, i_k) := P_k$, i.e.,
the last move added.
In order to make space for $j_k$, the machine $i_k$ should repel jobs.
\begin{description}
  \item[(small-all)] If $j_k$ is small, let $i_k$ repel all jobs.
\end{description}
In the case that $j_k$ is big, we need to be more careful.
It helps to imagine that the algorithm is a lazy one: It
repels jobs only if it is really necessary.

For $i\in\machines$ let $S_i(P_{\le k-1}, \sigma)$ be those small jobs $j$ which have $\sigma(j) = i$
and which are repelled by all other potential machines, i.e., $\Gamma(j)\setminus\{i\}$,
w.r.t. $P_{\le k-1}$.
The intuition behind $S_i(P_{\le k-1}, \sigma)$ is that we do not expect that 
$i$ can get rid of any of these jobs.

Next, define a threshold $W_0$ as the minimum
$W \ge 0$ such that the small jobs in $S_{i_k}(P_{\le k-1}, \sigma)$ and
all big jobs below this threshold are already too large to move $j_k$, i.e.,
\begin{equation*}
  p(S_{i_k}(L_{\le k-1}, \sigma)) + p(\{j\in\sigma^{-1}(i_k) : 1/2 < p_j \le W\}) + p_{j_k} > 11/6 \cdot \tau .
\end{equation*}
Furthermore, define $W_0 = \infty$ if no such $W$ exists.
In order to make $(j_k, i_k)$ valid,
it is necessary (although not always sufficient) to remove one
of the big jobs with size at most $W_0$.
Hence, we define,
\begin{description}
\item[(big-all)] if $j_k$ is big and $W_0 = \infty$, then let $i_k$ repel all jobs w.r.t $P_{\le k}$ and
\item[(big-big)] if $j_k$ is big and $W_0 < \infty$, then let $i_k$ repel $S_{i_k}(L_{\le k-1}, \sigma)$ and all jobs $j$ with $1/2 < p_j \le W_0$.
\end{description}
Note that repelling $S_{i_k}(P_{\le k-1}, \sigma)$ seems unnecessary, 
since those jobs do not have any machine to go to.
However, this definition simplifies the analysis.
It is also notable that the special case where $W_0 = 0$ is equivalent to
$p(S_{i_k}(L), \sigma) + p_{j_k} > 11/6 \cdot \tau$ and here the algorithm gives up making $(j_k, i_k)$ valid.
Finally, we want to highlight the following counter-intuitive (but intentional) aspect of the algorithm.
It might happen that some job of size greater
than $W_0$ is moved to $i_k$, but has to be moved again later in order to make $(j_k, i_k)$ valid.

\subsection{Analysis}
\begin{lemma}
  If the configuration LP is feasible for $\tau$ and there is a bad machine,
  the algorithm always finds a move to execute.
\end{lemma}
\begin{proof}
Suppose toward contradiction, there is a bad machine, no move in $P_{\le\ell}$ is valid and
no move can be added to $P_{\le\ell}$.
We will construct values $(z_j)_{j\in\jobs}$, $(y_i)_{i\in\machines}$ with the properties as in Lemma~\ref{lemma:condition-ra} and
thereby show that the configuration LP is infeasible.
During this proof, $\sigma$ and $P_{\le\ell}$ are constant and refer to the state at which the algorithm
is stuck. We omit $P_{\le\ell}$ when we say $i$ repels $j$.

Let $\tilde\jobs_i$ denote all jobs $j\in\sigma^{-1}(i)$
that are repelled by $i$.
We write $\tilde\jobs=\bigcup_{i\in\machines}\tilde\jobs_i$.
For every $j\in\jobs$ let
\begin{equation*}
  z_j = \begin{cases}
    \min\left\{\frac{p_j}{\tau}, \frac 5 6 \right\} &\text{if $j\in\tilde\jobs$ and} \\
    z_j = 0 &\text{otherwise}.
  \end{cases}
\end{equation*}
Let $y_i := 1$ if $i\in\machines$ repels all jobs and $y_i = z(\sigma^{-1}(i))$ otherwise.

Let $i\in\machines$ and
$C\in\mathcal C(i, \tau)$.
We need to show that $z(C) \le y_i$.
If $y_i = 1$ this follows immediately, because $z(C) \le p(C) / \tau \le 1$.
We assume w.l.o.g. that $i$ does not repel all jobs 
and thus $y_i = z(\sigma^{-1}(i))$.
In particular, $i$ does not repel small jobs that are on other machines.
This means that $z_j = 0$ for every small job $j\in C\setminus\sigma^{-1}(j)$.
Otherwise, $(j, i)$ could be added to $P_{\le\ell}$.
If there are no big jobs in $C$, we therefore 
get $z(C) \le z(\sigma^{-1}(i)) = y_i$.
Clearly, there can be at most one big job $j_B\in C$,
since such a job has $p_{j_B} > 1/2 \cdot \tau$ and $C$ cannot have a load greater than $\tau$.
If $z_{j_B} = 0$ or $\sigma(j_B) = i$ the argument above still holds.

We recap: The only interesting case is when $y_i = z(\sigma^{-1}(i))$, there is exactly one big job $j_B \in C\setminus\sigma^{-1}(i)$, and $z_{j_B} = \min\{p_{j_B}/\tau, 5/6\}$.
\begin{description}
\item[Case 1: $i$ repels $j_B$.]
Since $i$ is not the target of a small job move and it is good (otherwise it would repel all jobs),
there must be a big job move that causes $i$ to repel $j_B$.
In other words, there is a move $(j_k, i)=P_k$ such that $j_B\le W_0$, where $W_0$ is as in the definition of
repelled jobs for $P_{\le k}$.
Recall that $W_0 < \infty$ is the minimal $W$ with
\begin{equation*}
  p(S_i(P_{\le k - 1}, \sigma)) + p(\{j\in\sigma^{-1}(i) : 1/2 < p_j \le W\}) + p_{j_k} > 11/6 \cdot \tau .
\end{equation*}
Since $W_0 \ge p_{j_B}$ and it is minimal,
there must be a big job $j_B'\in\sigma^{-1}(i)$ with
$p_{j_B'} = W_0 \ge p_{j_B}$ and $j_B'$ is also repelled by $i$ (because $p_{j'_B} \le W_0$).
We get
\begin{equation*}
z(C) \le z(C\setminus \{j_B\}) + z_{j_B} \le z(\sigma^{-1}(i)\setminus\{j_B'\}) + z_{j_B'} = z(\sigma^{-1}(i)) = y_i.
\end{equation*}
\item[Case 2: $i$ does not repel $j_B$.]
Since $(j_B, i)$ cannot be added to $P$, it must already be in $P$. Let $P_k = (j_B, i)$
and $W_0$ as in the definition of repelled edges w.r.t. $P_{\le k}$.
Then
\begin{equation*}
  p(S_i(P_{\le k-1},\sigma)) + p(\{j\in\sigma^{-1}(i) : 1/2 < p_j \le W_0\}) + p_{j_B} > 11/6 \cdot \tau .
\end{equation*}
If $W_0\ge 5/6$, then there is some $j'_B\in\sigma^{-1}(i)$ with $p_{j'_B} = W_0 \ge 5/6$.
Similar to the previous case, it follows that
\begin{equation*}
z(C) \le z(C\setminus \{j_B\}) + z_{j_B} \le z(\sigma^{-1}(i)\setminus\{j'_B\}) + 5/6 = z(\sigma^{-1}(i)) = y_i.
\end{equation*}
If $W_0\le 5/6$, then all of the considered jobs have $z_j = p_j/\tau$, i.e.,
\begin{align*}
  y_i &= z(\sigma^{-1}(i)) \\
  &\ge z(S_i(P_{\le k-1},\sigma)) + z(\{j\in\sigma^{-1}(i) : 1/2 < p_j \le W_0\}) \\
  &= p(S_i(P_{\le k-1},\sigma))/\tau + p(\{j\in\sigma^{-1}(i) : 1/2 < p_j \le W_0\})/\tau \\
  &> 11/6 - p_{j_B}/\tau \ge 5/6 + (\tau-p_{j_B})/\tau \ge z(C) .
\end{align*}
The last inequality holds, because the $z_{j_B}$ is at most $5/6$ and the volume of the small jobs in $C$
(in particular, their value) is at most $(\tau-p_{j_B})/\tau$.
\end{description}
It remains to show that $\sum_{j\in\jobs} z_j > \sum_{i\in\machines} y_i$. We prove that, with amortization, good machines
satisfy $z(\sigma^{-1}(i)) \ge y_i$ and on bad machines
strict inequality holds.
Let $i$ be a bad machine.
Then $i$ repels all jobs (in particular, those in $\sigma^{-1}(i)$).
Hence,
\begin{equation*}
  z(\sigma^{-1}(i)) \ge 5/6 \cdot p(\sigma^{-1}(i)) / \tau > 55/36 > 1 = y_i .
\end{equation*}
For good machines that do not repel all jobs, equality holds by definition.
We will partition those good machines that do repel all jobs
into those $i\in\machines$  which have $(j_S, i) \in P_{\le\ell}$ for a small
job $j_S$ and those that do not.
\begin{lemma}
  \label{min-max-ratio}
  At least half of the machines $i$ that repel all jobs are target of a small job, i.e.,
  $(j_S, i)\in P_{\le\ell}$ for some small job $j_S$.
\end{lemma}
We argue that whenever a move $(j_B, i)=P_k$ of a big job $j_B$ is added, it is not valid, and
$W_0 = \infty$ in the definition of repelled jobs w.r.t. $P_{\le k}$, then there
is some small job move $(j_S, i')$ that can be added to $P$. Since the algorithm
prefers small job moves over big ones, the next move after $P_k$ will necessarily be a small job move.
Since no two small job moves can be added for the same target, the lemma follows.
If $W_0 = \infty$, this means that
\begin{equation*}
  p(S_i(L_{\le k-1},\sigma)) + p(\{j\in\sigma^{-1}(i) : p_j > 1/2\}) + p_{j_B} \le 11/6\cdot\tau .
\end{equation*}
Since the move $(j_B, i)$ is not valid, however, we also have that
\begin{equation*}
  p(\sigma^{-1}(i)) + p_{j_B} > 11/6\cdot\tau .
\end{equation*}
This implies that there must be a small job $j_S\in\sigma^{-1}(i)\setminus S_i(L_{\le k-1}\sigma)$.
In particular, there exists some $i'\in\Gamma(j_S)\setminus\{i\}$ by which $j_S$ is not repelled.
It is also not repelled by $i'$ w.r.t. $P_{\le k}$, since $P_k$ only adds
repelled jobs for $i$. Therefore, $(j_S, i')$ is a candidate for the next move to be added after $P_{\le k}$.
This concludes the proof for the lemma.

Let $i$ be a machine that repels all jobs, but is not target of a small job.
Then there is a big job $j_B$ with $(j_B, i) \in P_{\le\ell}$ and this move is not valid. 
Either there is a job $j\in\sigma^{-1}(i)$ with $z_j = 5/6$ or $z_j = p_j / \tau$ for all $j\in\sigma^{-1}(i)$. Thus,
\begin{equation*}
  z(\sigma^{-1}(i)) \ge \min\left\{\frac 5 6,\ \frac{p(\sigma^{-1}(i))}{\tau}\right\}
  \ge \min\left\{\frac 5 6,\ \frac{11/6 - p_{j_B}}{\tau}\right\}
  \ge \frac 5 6 = y_i - \frac 1 6 .
\end{equation*}
Next, let $i$ be a machine such that there exists a small job $j_S$ with $(j_S, i) \in P_{\le\ell}$.
This move is also not valid.
In the following, we distinguish between the cases where $\sigma^{-1}(i)$ has no job
$j$ with $z_j = 5/6$, one such job, or at least two. Note that all jobs have $p_j \le \tau$.
\begin{multline*}
  z(\sigma^{-1}(i)) \ge \min\left\{\frac{p(\sigma^{-1}(i))}{\tau},\ \frac{p(\sigma^{-1}(i)) - \tau}{\tau} + \frac 5 6,\ \frac{10}{6}\right\} \\
  \ge \min\left\{\frac{11}{6} - \frac{p_{j_S}}{\tau} - 1 + \frac 5 6,\ \frac{10}{6} \right\}
  \ge \frac 7 6 = y_i + \frac 1 6 .
\end{multline*}
Because of Lemma~\ref{min-max-ratio}, we can amortize and get
\begin{equation*}
  \sum_{j\in\jobs} z_j = \sum_{i\in\machines}z(\sigma^{-1}(i)) > \sum_{i\in\machines} y_i . \qedhere
\end{equation*}
\end{proof}
\begin{lemma}
  The algorithm terminates.
\end{lemma}
\begin{proof}
Consider two consecutive iterations of the main loop right before a move is executed and
the list of moves $P$ is deleted.
Let $\sigma$, $P_{\le\ell}$ be the allocation and list of pending moves in the former iteration
and $\sigma'$, $P'_{\le\ell'}$ in the latter.
Let $b$ and $b'$ be the number of bad machines in $\sigma$ and $\sigma'$.
Recall, $R(P_{\le k}, \sigma)\subseteq \jobs\times\machines$ is the set of all $i, j$
where $j$ is repelled by $i$ w.r.t. $P_{\le k}$.
Further, let $\tilde\jobs(P_{\le k}, \sigma)$ denote all jobs $j$
that are repelled by $\sigma(j)$ w.r.t. $P_{\le k}$.
For each prefix of $P$ (and of $P'$)
we define a potential function
\begin{equation*}
  \Phi(P_{\le k}, \sigma) = (p_{j_k}, j_k, i_k, |\jobs\times\machines| - |R(P_{\le k}, \sigma)|, |\tilde\jobs(P_{\le k}, \sigma)|)
\end{equation*}
We claim that the vector
\begin{equation*}
  (b', |\tilde\jobs(P'_{\le 0}, \sigma')|, \Phi(P'_{\le 0}, \sigma'), \dotsc, \Phi(P'_{\le \ell'}, \sigma'), -1)
\end{equation*}
is lexicographically smaller than
\begin{equation*}
  (b, |\tilde\jobs(P_{\le 0}, \sigma)|, \Phi(P_{\le 0}, \sigma), \dotsc, \Phi(P_{\le \ell}, \sigma), -1) .
\end{equation*}
Note that the length of the vector is bounded by $5\cdot |\machines|\cdot |\jobs| + 3$,
since no move appears twice in the list and every component can have at most
$|\jobs| \cdot |\machines| + 1$ different values. Thus, the number of possible vectors
is finite and hence the algorithm terminates.

Recall that the algorithm never turns a good machine bad, which means $b'\le b$.
If $b' < b$, we are done. Likewise, if $b=b'$ and some job is moved
from a bad machine to a good machine,
then $|\tilde\jobs(P'_{\le 0}, \sigma')| < |\tilde\jobs(P_{\le 0}, \sigma)|$
and again the first vector is lexicographically smaller. We can
therefore focus on the case $b' = b$,
$\tilde\jobs(P'_{\le 0}, \sigma') = \tilde\jobs(P_{\le 0}, \sigma)$,
and $\sigma(j) = \sigma'(j)$ for all $j\in\tilde\jobs(P_{\le 0}, \sigma)$.
The rest of the argument is by induction. Let $k\le\min\{\ell-1,\ell'\}$ and assume that
\begin{enumerate}
  \item $P'_{\le k-1} = P_{\le k-1}$,
  \item $R(P'_{\le k-1}, \sigma') = R(P_{\le k-1}, \sigma)$.
  \item $\tilde\jobs(P'_{\le k-1}, \sigma') = \tilde\jobs(P_{\le k-1}, \sigma)$;
    $\sigma'(j) = \sigma(j)$ for all $j\in\tilde\jobs(P_{\le k-1}, \sigma)$,
\end{enumerate}
We will show that $\Phi(L'_{\le k}) \le \Phi(L_{\le k})$ (lexicographically) and 
if equality holds, then (1), (2), and (3) also hold for $k$.
This implies the lexicographical decrease:
If $\ell' < \ell$ it follows easily. This is because the prefix of the first
vector ending in $\Phi(P'_{\le \ell'})$ is lexicographically not bigger than
the prefix of the second vector ending in $\Phi(P_{\le \ell'})$. Furthermore,
the next component is $-1$ in
the first vector, but something non-negative in the other.
Now consider the case $\ell'\ge \ell$.
Let $(j_\ell, i_\ell)=P_\ell$ be the
move that was executed. Then $\sigma'(j_\ell) = i_\ell \neq \sigma(j_\ell)$.
Furthermore, $j_\ell$ is repelled by $\sigma(j_\ell)$ w.r.t. $L_{\le \ell-1}$.
Hence, (2) cannot hold with $k-1=\ell-1$ and thus
we cannot have $\Phi(L'_{\le k}) = \Phi(L_{\le k})$ for all $k$.

The approach for the induction is to show that if $\Phi(L'_{\le k}, \sigma')$ is not smaller
than $\Phi(L_{\le k}, \sigma)$, $(j_k, i_k) = P_k$ also has to be selected
as $P'_k$. In that case, no job can have been moved to $i_k$, if it is
repelled by $i_k$ w.r.t. $L_{\le k}$. From the way they are chosen,
this implies the jobs which $i_k$ repels as a consequence of $P_k$ are also repelled in
the rules for $P'_k$.
If they do not increase, the number of jobs $j$ with $\sigma(j)$ that are repelled by $i_k$ 
cannot increase. Let us now formalize this argument.

Notice that $(j_\ell, i_\ell) = P_\ell$ is the move that
was executed, i.e., $\sigma'(j_\ell) = i_\ell$
 and by construction of $P_{\le\ell}$,
$j_\ell$ is not repelled by $i_\ell$ w.r.t. $P_{\le \ell-1}$
(in particular, not w.r.t. $P_{\le k}$ $(*)$).
Let $(j_k, i_k) = P_k$. By (1) we have that $(j_k, i_k)\notin P_{\le k-1} = P'_{\le k-1}$. Since $j_k$ is repelled by $\sigma(j_k)$ w.r.t. $L_{\le k}$, by (2) we have that
$\sigma'(j_k) = \sigma(j_k)$. By (3) it is not repelled by $i_k$ w.r.t. $L'_{\le k-1}$.
Therefore, $(j_k, i_k)$ is was a candidate for $P'_k$.
Either this or a move $(j, i)$, where $(p_j, j, i)$ is lexicographically smaller
than $(p_{j_k}, j_k, i_k)$ is chosen. In the latter case we have
$\Phi(L'_{\le k}) < \Phi(L_{\le k})$. Hence, assume that $P'_k = (j_k, i_k)$.
This means (1) holds for $k$.
Note that since (2) and (3) hold for $k-1$, we only have to check the consequences of $P'_k$ and $P_k$.
In other words, we have to check whether the rules for move $P'_k$ imply the same repelled jobs as $P_k$
and whether some job repelled due to this rule has been moved.

If $j_k$ is small, then $i_k$ repels all jobs w.r.t. $L'_{\le k}$ and $L_{\le k}$
and therefore (3) also holds for $k$. Job $j_\ell$ cannot have been moved to
$i_k$ (see $(*)$). If it was moved away from $i_k$, then
$\tilde\jobs(P'_{\le k}, \sigma') \subsetneq \tilde\jobs(P_{\le k-1}, \sigma)$.
Otherwise, equality holds and no job was moved in this set, i.e., (2) holds for $k$.

Now assume $j_k$ is big.
First, we argue that $S'_{i_k}(L'_{\le k-1}, \sigma') = S_{i_k}(L_{\le k-1}, \sigma)$.
Recall, $S_{i_k}(L_{\le k-1}, \sigma)$ are the small jobs $j$ 
with $\sigma(j) = i_k$ and $j$ is repelled by
all machines in $\Gamma(j)\setminus\{i_k\}$ w.r.t. $L_{\le k}$.
Let $j\in S'_{i_k}(L'_{\le k-1}, \sigma')$.
Assume toward contradiction that
$\sigma(j)\neq i_k = \sigma'(j)$. Then $j=j_\ell$ and $i_k=i_\ell$.
However, $\sigma(j_\ell)$ does not repel $j_\ell$ w.r.t. $P_{\le k-1}$. Otherwise
$(j_\ell, i_\ell)$ would have been chosen instead of $(j_k, i_k)$ as $P_k$.
By (3) $\sigma(j_\ell)$ also does not repel $j_\ell$ w.r.t. $P'_{\le k-1}$.
Hence, $j\notin S'_{i_k}(L'_{\le k-1},\sigma')$, a contradiction.
Consequently, $\sigma(j) = i_k$. By (3) it follows that $j\in S_{i_k}(L_{\le k-1},\sigma)$.
Let $j\in S_{i_k}(L_{\le k-1},\sigma)$. Since it is repelled by all potential machines w.r.t.
$P_{\le k-1}$, there cannot be a move for $j$ in a later layer. In particular $j_\ell$ cannot
be $j$. This means $\sigma'(j) = \sigma(j) = i_k$ and by (3)
$j\in S_{i_k}(L_{\le k-1})$.
We conclude that $S_{i_k}(L_{\le k-1})=S'_{i_k}(L'_{\le k-1})$.
Let $W_0$ be the minimal the minimal $W\ge 0$ with
\begin{equation*}
  \underbrace{p(S_{i_k}(L_{\le k-1}, \sigma))}_{=p(S'_{i_k}(L'_{\le k-1},\sigma'))} + p(\{j\in\sigma^{-1}(i) : 1/2 < p_j \le W\}) + p_{j_k} > 11/6\cdot \tau ,
\end{equation*}
or $\infty$ if no such $W$ exists.
Since all jobs $j$ with $1/2 < p_j \le W_0$ are repelled by $i_k$ w.r.t. $P_{\le k}$,
it follows by $(*)$ that
\begin{equation*}
  p(\{j\in\sigma^{\prime-1}(i) : 1/2 < p_j \le W\}) \le
  p(\{j\in\sigma^{-1}(i) : 1/2 < p_j \le W\})
\end{equation*}
It follows that $W'_0$, the minimal $W\ge 0$ with
\begin{equation*}
  p(S'_{i_k}(L'_{\le k-1},\sigma')) + p(\{j\in\sigma^{\prime-1}(i) : 1/2 < p_j \le W\}) + p_{j_k} > 11/6\cdot \tau ,
\end{equation*}
is at least as big as $W_0$. In particular, if $W_0 = \infty$, then $i_k$ repels all jobs w.r.t. $P_{\le k}$
and w.r.t. $P'_{\le k}$.
Otherwise, by definition $i_k$ repels all jobs in $S_i(L_{\le k-1},\sigma)$ and all
jobs $j$ with $1/2 < p_j \le W_0$ w.r.t. $L_{\le k}$.
By $W'_0 \ge W_0$ these jobs are also repelled by $i_k$ w.r.t. $L'_{\le k}$.
Hence $R(P'_{\le k}, \sigma') \supseteq R(P_{\le k}, \sigma)$.
If equality holds then because of $(*)$, 
$\tilde\jobs(P'_{\le k}, \sigma') \supseteq \tilde\jobs(P_{\le k}, \sigma)$.
If equality also holds here, then none of these jobs are moved and (1), (2), and (3) hold.
If equality does not hold at some point, $\Phi(L'_{\le k-1},\sigma') < \Phi(L_{\le k-1},\sigma)$.
\end{proof}

\begin{theorem}\label{th-ra}
  The integrality gap of the configuration LP for \textsc{Restricted Assignment}
  is at most $11/6$.
\end{theorem}

\section{Quasi-polynomial time algorithm}
The previous running time bound is clearly exponential.
In this section, we will improve the running time to
$n^{O(1/\epsilon \cdot \log(n))}$ for an approximation rate of
$11/6 + 2\epsilon$,
where $n = |\machines| + |\jobs|$ and $\epsilon > 0$ can be chosen arbitrarily. Note that by scaling $\epsilon$, the coefficient of $2$ can be removed.
\subsection{Algorithm}
Our approach for turning the running time quasi-polynomial 
is to combine the two algorithms presented in the previous sections.
Instead of adding only one candidate move at a time like in the exponential time algorithm,
we add them all. The set of these moves is called a layer. After a layer
is added, re-evaluate the repelled jobs and again construct a layer of all sensible moves. 

This simple approach described has some major issues that we have to handle carefully
using more sophisticated techniques as described in the following.
First, however, we make some technical preparations. 
We will call a machine $i$ \emph{bad}, if $p(\sigma^{-1}(i)) > (11/6 + 2\epsilon)\tau$ and \emph{good},
otherwise.
As before, we call jobs $j$ small, if $p_j \le 1/2 \cdot\tau$ and big otherwise.
Further, we distinguish big jobs into medium, which have $p_j\le 5/6\cdot\tau$, and huge,
which have $p_j > 5/6\cdot\tau$.
Unlike in the previous algorithms we establish the invariant
that at all times the current allocation
assigns at most one huge job to each machine.
An initial allocation that satisfies this can easily be found
via bipartite matching with the huge jobs
on one side and the machines on the other.
It is maintained by the following definition of a valid huge job move.
\begin{definition}[Valid huge move]
  Let $j\in\jobs, i\in\Gamma(j)\setminus\{\sigma(j)\}$ with $p_j > 5/6\cdot\tau$. $(j, i)$ is a valid move,
  if $p(\sigma^{-1}(i)) + p_j \le (11/6 + 2\epsilon)\tau$ and there is no huge job
  in $\sigma^{-1}(i)$.
\end{definition}
For the remaining jobs we define the moves as follows.
\begin{definition}[Valid non-huge move]
  Let $j\in\jobs, i\in\Gamma(j)\setminus\{\sigma(j)\}$ with $p_j \le 5/6\cdot \tau$. $(j, i)$ is a valid move, if
\begin{enumerate}
  \item $p(\sigma^{-1}(i)) + p_j \le (11/6 + 2\epsilon)\tau$ and $\sigma^{-1}(i)$ contains no huge job, or
  \item $p(\{j'\in \sigma^{-1}(i) : p_j \le 5/6\cdot \tau\}) + p_j \le (5/6 + 2\epsilon)\tau$ and $\sigma^{-1}(i)$ contains one huge job.\label{small-huge}
\end{enumerate}
\end{definition}
Each valid move $(j, i)$ satisfies $p(\sigma^{-1}(i)) + p_j \le (11/6 + 2\epsilon)\tau$,
i.e., each good machine stays good.
(\ref{small-huge}) needs further elaboration.
One could falsely assume that this establishes
an invariant which says the non-huge load is at most $(5/6+2\epsilon)\cdot\tau$ on a machine
with a huge job.
This would be a marvelous invariant, if it could be guaranteed.
However, a valid huge move can break this property.
Therefore, (\ref{small-huge}) only gives something weaker. It's purpose is that
when a machine has a huge job and a low non-huge load, then it will
stay this way for as long as the huge job remains on the machine.
This is to keep edges in the leap graph intact, a technique that will be elaborated later.
\paragraph*{Layers}
We will operate in layers $L_1, \dotsc, L_{\ell}$ like in the first algorithm.
These, however, will not only contain machines, but also fine grained moves
like in the second algorithm.
Again, we define a binary relation $R(L_{\le k}, \sigma)\subseteq \jobs\times\machines$,
which states that $i$ repels $j$ w.r.t. $L_{\le k}$, if $(j, i)\in R(L_{\le k}, \sigma)$.
In the previous algorithms, 
the local search is mostly stateless, i.e., it searches for an
improvement of $\sigma$ without remembering anything from the past.
Here we make a small exception. We maintain an order $\pi$ on $\jobs\times\machines$.
When the algorithm adds certain critical moves, they are moved to the front of $\pi$.
This will hint at the algorithm to do the same in the next iteration.
It helps to argue about the running time, since the layers created this way
are more consistent throughout the iterations.
Finally, layers come in different forms.
They can be leap layers, critical layers, small layers, or non-critical layers.
This will become more clear in the actual description of the algorithm.
\paragraph*{Leaps}
A straight-forward example shows that if we are using only simple moves like in the previous
two algorithms, the number of layers needs to grow linearly.
This would be a problem for obtaining a quasi-polynomial running time.
\begin{figure}
\centering
\begin{tikzpicture}
  \draw[very thick] (2, 1.5) -- (2, -0.1) -- (2.5, -0.1) -- (2.5, 1.5);
  \draw[thick] (2.1, 0) rectangle (2.4, 1);
  \draw[thick] (2.1, 1) rectangle (2.4, 2);
  \draw[->, thick] (2.2, 1.5) -- (5.4, 2) node[above, yshift=0.2cm, pos=0.5] {can move here};
  \draw[very thick] (5.5, 3) -- (5.5, 1.4) -- (6, 1.4) -- (6, 3);
  \draw[thick] (5.6, 1.5) rectangle (5.9, 2.5);
  \draw[->, thick] (5.7, 2) -- (6.9, 2);
  \node at (7.2, 2) {\dots};
  \draw[->, thick] (7.5, 2) -- (8.4, 2);
  \draw[very thick] (3+5.5, 3) -- (3+5.5, 1.4) -- (3+6, 1.4) -- (3+6, 3);
  \draw[thick] (3+5.6, 1.5) rectangle (3+5.9, 2.5);
  \draw[->, thick] (3+5.7, 2) -- (5+5.4, 2);
  \draw[very thick] (5+5.5, 3) -- (5+5.5, 1.4) -- (5+6, 1.4) -- (5+6, 3);

  \draw[->, thick] (2.2, 0.5) -- (5.4, 0);
  \draw[very thick] (5.5, 1) -- (5.5, -0.6) -- (6, -0.6) -- (6, 1);
  \draw[thick] (5.6, -0.5) rectangle (5.9, 0.5);
  \draw[->, thick] (5.7, 0) -- (6.9, 0);
  \node at (7.2, 0) {\dots};
  \draw[->, thick] (7.5, 0) -- (8.4, 0);
  \draw[very thick] (3+5.5, 1) -- (3+5.5, -0.6) -- (3+6, -0.6) -- (3+6, 1);
  \draw[thick] (3+5.6, -0.5) rectangle (3+5.9, 0.5);
  \draw[->, thick] (3+5.7, 0) -- (5+5.4, 0);
  \draw[very thick] (5+5.5, 1) -- (5+5.5, -0.6) -- (5+6, -0.6) -- (5+6, 1);
\end{tikzpicture}
\caption{Example for linear number of layers}
\label{fig:lin}
\end{figure}

In the example (see Fig.~\ref{fig:lin}) the leftmost machine has two jobs $j_1, j'_1$
of size $\tau$ assigned to it,
which make the machine bad. The jobs each have a chain of machines connected to it:
$j_1$ can go to a machine $i_1\in\Gamma(j_1)$. On $i_1$ there is another job $j_2$
of size $\tau$ which can go to a machine $i_2\in\Gamma(j_2)$, etc.
At some point this chain ends with an empty machine. The same construction is made
for $j'_1$.
In order to make the bad machine good, either the top chain of jobs or the bottom chain
has to be traversed. Hence, it seems like
the number of layers would be roughly half the jobs or machines.
It turns out that this problem only occurs with huge jobs
and we will carefully circumvent it.

The leap technique is intended for moving such a chain of huge jobs at once.
For an easy description we construct a directed bipartite graph $G(\sigma) = (V, E(\sigma))$
where $V = B\cup\machines$, i.e., the vertices are big jobs $B$ and the machines.
There is an edge $(j_B, i)\in E(\sigma)$ if $i\in\Gamma(j_B)\setminus\{\sigma(j_B)\}$ and
\begin{equation*}
  p(\{j\in\sigma^{-1}(i) : p_{j} \le 5/6\}) + p_{j_B} \le (11/6 + 2\epsilon) \tau ,
\end{equation*}
i.e., if there is no huge job in $\sigma^{-1}(i)$ the move $(j_B, i)$ is
valid. Furthermore, we let $(i, j_H)\in E(\sigma)$, if $j_H$ is huge and $i = \sigma(j_H)$.

Suppose that there is some path $j_1, i_1, j_2, i_2,\dotsc, j_k, i_k$
in $G$, where no huge job is assigned to $i_k$.
Then we can move $j_1$ to $i_1$,
$j_2$ from $i_1$ to $i_2$, $j_3$ from $i_2$ to $i_3$, etc.
We will call this a \emph{leap}.

The general theme for using this graph is the following.
A big job $j_B$ is repelled by $\sigma(j_B)$. Then all machines that
are reachable by some path from $j_B$ should repel their huge jobs as well. When
one of them is removed, we can instantly free $i$ from $j_B$.
This way we are not going to put all moves of the path sequentially into the layers and avoid
making it unnecessarily long.
\begin{definition}[Valid leap]
  Let $j_1,i_1,\dotsc,j_r,i_r$ be a path in the leap graph.
  It is called valid leap, if $(j_r,i_r)$ is a valid move.
\end{definition}
By definition of the leap graph and the fact that every machine has at most one huge job, 
for a valid leap the following moves are all valid if executed in reverse order, i.e.,
$(j_r, i_r), (j_{r-1}, i_{r-1}),\dots, (j_1,i_1)$.

Finally, we define a graph $G(L_{\le k}, \sigma)$ which has all edges from $G(\sigma)$
of the form $(i, j_H)$, but only the edges $(j_B, i)$ where $i$ does not repel $j_B$ w.r.t.
$L_{\le k}$. The definition of repelled edges will be given later.
\paragraph{Description of the algorithm}
We are now ready to state the algorithm (see Alg.~\ref{alg:ra2}).
It starts with an allocation $\sigma$ with the property that each machine
has at most one huge job assigned to it. The allocation of huge jobs
can be found using bipartite matching and the remaining jobs are assigned arbitrarily.
We initialize $\pi$ as an arbitrary permutation of $\jobs\times\machines$.
Until all machines are good the algorithm searches for valid moves or leaps that
improve $\sigma$. For this purpose we build layers. The layers are alternating
between leap layers, critical layers, small layers, and non-critical layers: 
Layer $L_{4k+1}$ is always a leap layer; $L_{4k+2}$ is a critical layer;
$L_{4k+3}$ is a small layer; $L_{4k+3}$ is a non-critical layer.
A leap-layer $L_{\ell+1}$ consists of the machines that are reachable in the leap graph
$G(L_{\le\ell},\sigma)$ by a job that is repelled by its current machine w.r.t. $L_{\le\ell}$.
For the critical layer $L_{\ell+2}$ and non-critical layer $L_{\ell+4}$,
we select all $(j_B, i)$ where $j_B$ is a big job repelled by
$\sigma(j_B)$, but not by $i$ w.r.t. $L_{\le \ell+1}$. A subset of these
is taken in $L_{\ell+2}$.
We will define below precisely how they are chosen, but they depend on $\pi$
giving priority to the moves in the front.
After they are selected, the critical moves are pushed to the front of $\pi$.
All moves $(j_S, i)$, where $j_S$ is repelled by $\sigma(j_S)$, but not by $i$ w.r.t. $L_{\le \ell+2}$
are put in $L_{\ell+3}$.
Finally, the previously considered big job moves $(j_B, i)$ which were not taken in $L_{\ell+2}$
and where $i$ still does not repel $j_B$ (now w.r.t. $L_{\le\ell+3}$) are taken in 
the non-critical layer $L_{\ell+4}$.
If at any point a valid leap or move is found, it is executed and the structure of
layers is reset. Note that the meaning of
the \emph{continue} statement in the pseudo-code is to
jump to the next iteration of the while loop.

\begin{algorithm}
\caption{Constructive algorithm for \textsc{Restricted Assignment}}
\label{alg:ra2}

\begin{algorithmic}
\STATE{let $\sigma$ be an allocation with at most one huge job on each machine}
\STATE{let $\pi$ be an arbitrary order on $\jobs\times\machines$}
\STATE{$\ell \gets 0$}
\WHILE{there is a bad machine}
\IF{$\ell \ge 4\lceil\log_{1+\epsilon}(4|\machines|)\rceil = O(1/\epsilon \cdot \log(|\machines|))$}
  \RETURN "err"
\ENDIF
\STATE{let $L^L_{\mathrm{new}}$ be the set of machines reachable in the
  leap graph $G(L_{\le\ell},\sigma)$
  by a big job $j$ repelled by $\sigma(j)$ w.r.t. $L_{\le\ell}$}
\STATE{$L_{\ell+1}\gets (L^L_{\mathrm{new}}, \text{\textsc{leap}})$}
\IF{there is a machine $i$ in $L_{\ell+1}$ with no huge job in $\sigma^{-1}(i)$}
  \STATE{let $j_1,i_1,\dotsc,j_r,i_r=i$ be a path in $G(L_{\le\ell},\sigma)$, where $j_1$ is repelled by $\sigma^{-1}(j_r)$ w.r.t. $L_{\le \ell}$}
  \STATE{$\sigma(j_r)\gets i_r,\dotsc, \sigma(j_1)\gets i_1$}
  \STATE{delete $L_{1},L_{2},\dotsc,L_{\ell+1}$}
  \STATE{$\ell \gets 0$}
  \STATE{continue}
\ENDIF
\STATE{let $L^B_{\mathrm{new}}$ be the set of all $(j_B, i)$, $j_B\in\jobs$ big and $i\in\Gamma(j_B)$, with
  $j_B$ repelled by $\sigma(j_B)$ and not by $i$ w.r.t. $L_{\le \ell+1}$}
\STATE{$L^C_{\mathrm{new}}\gets \mathrm{CriticalMoves}(L^B_{\mathrm{new}}, \sigma, L_{\le\ell+1}, \pi)$}
\STATE{$L_{\ell+2}\gets (L^C_{\mathrm{new}}, \text{\textsc{critical}})$}
\STATE{move $L^C_{\mathrm{new}}$ to the front of $\pi$ (keeping their pairwise order)}
\IF{there exists a valid move $(j, i)$ in $L_{\ell+2}$}
  \STATE{$\sigma(j) \leftarrow i$; delete $L_{1},L_{2},\dotsc,L_{\ell+2}$; $\ell \gets 0$}
  \STATE{continue}
\ENDIF
\STATE{let $L^S_{\mathrm{new}}$ be the set of all $(j_S, i)$, $j_S\in\jobs$ small and $i\in\Gamma(j_S)$, with
  $j_S$ repelled by $\sigma(j_S)$ and not by $i$ w.r.t. $L_{\le \ell+2}$}
\STATE{$L_{\ell+3}\gets (L^S_{\mathrm{new}}, \text{\textsc{small}})$}
\IF{there exists a valid move $(j, i)$ in $L_{\ell+3}$}
  \STATE{$\sigma(j) \leftarrow i$; delete $L_{1},L_{2},\dotsc,L_{\ell+3}$; $\ell \gets 0$}
  \STATE{continue}
\ENDIF
\STATE{let $L^{NC}_{\mathrm{new}}$ be the set of all $(j, i)\in L^B_{\mathrm{new}}\setminus L^C_{\mathrm{new}}$ where $i$ does not repel $j$ w.r.t. $L_{\le\ell+3}$}
\STATE{$L_{\ell+4}\gets (L^{NC}_{\mathrm{new}}, \text{\textsc{non-critical}})$}
\IF{there exists a valid move $(j, i)$ in $L_{\ell+4}$}
  \STATE{$\sigma(j) \leftarrow i$; delete $L_{1},L_{2},\dotsc,L_{\ell+4}$; $\ell \gets 0$}
  \STATE{continue}
\ENDIF
\STATE{$\ell\gets\ell+4$}
\ENDWHILE
\RETURN $\sigma$
\end{algorithmic}
\end{algorithm}

\paragraph*{Repelled jobs}
We define the repelled jobs of each machine inductively w.r.t. $L_{\le k}$, $k=0,1,\dotsc,\ell$.
\begin{description}
  \item[(initialization)] Let the bad machine repel every job w.r.t. $L_{\le 0}$.
  \item[(monotonicity)] If $i$ repels $j$ w.r.t. $L_{\le k}$,
    then let $i$ repel $j$ also w.r.t. $L_{\le k + 1}$.
\end{description}
The remaining rules regard $k > 0$ and we define repelled jobs for a layer $L_k$.
A layer $L_k$ may be a leap layer, a critical layer, a small layer, or a non-critical layer.
In the first case, $L_k$ contains a set of machines reachable in the leap graph. We define:
\begin{description}
  \item[(leap)] If $L_k$ is a leap layer, let every machine $i$ in $L_k$
    repel all big jobs that are adjacent to $i$ in the leap graph $G(\sigma)$%
---that is, all huge jobs in $\sigma^{-1}(i)$ and all big jobs $j_B$ with
$i\in\Gamma(j_B)$ and
\begin{equation*}
  p(\{j'\in\sigma^{-1}(i) : p_{j'}\le 5/6\}) + p_{j_B} \le (11/6+\epsilon)\tau .
\end{equation*}
\end{description}
\begin{description}
  \item[(critical)] If $L_k$ is a critical layer, for every move
    $(j, i)$ in $L_k$ let $i$ repel all jobs.
\end{description}
\begin{description}
  \item[(small)] If $L_k$ is a small layer, for every move
    $(j, i)$ in $L_k$ let $i$ repel all jobs.
\end{description}
Now assume that $L_k$ is a non-critical-layer and consider a move $(j, i)$ in $L_k$.
In the non-critical case the algorithm is lazy:
It repels jobs only if it is really necessary.
We first identify a set of small jobs that is unlikely to be moved.
For $i\in\machines$ define $S_i(L_{\le k-1}, \sigma)$ to be the small jobs
$j\in\sigma^{-1}(i)$
which are repelled by all machines in $\Gamma(j)\setminus\{i\}$.

Next, define a threshold $W_0$ as the minimum
$W \ge 0$ such that the small jobs in $S_{i}(L_{\le k-1}, \sigma)$ and
all big jobs below this threshold
are already too large to add $j$, i.e.,
\begin{equation*}
  p\left(\left\{j'\in\sigma^{-1}(i) : \frac 1 2 < p_{j'} \le W\right\}\right)
+ p(S_{i}(L_{\le k-1}, \sigma)) + p_{j}
  > \left(\frac{11}{6} + 2\epsilon\right) \tau .
\end{equation*}
It will follow from the selection of critical moves that such a $W_0$ always exists
and, moreover, $W_0 \le 5/6$.
When none of the jobs in $S_{i}(L_{\le k-1}, \sigma)$ can be removed,
it is necessary (although not always sufficient) to remove one
of the big jobs with size at most $W_0$ in order to make $(j, i)$ valid.
Hence, we define,
\begin{description}
\item[(non-critical)] if $L_k$ is a non-critical layer then for every move $(j, i)$ let $i$
repel all jobs $j$ with $1/2 < p_j\le W_0$ (where $W_0$ is defined as above) and all jobs in $S_{i}(L_{\le k-1},\sigma)$.
\end{description}
It is notable that the corner case where $W_0 = 0$ is equivalent to
\begin{equation*}
  p(S_{i}(L_{\le k-1},\sigma)) + p_{j} > (11/6 + 2\epsilon) \tau
\end{equation*}
and here the algorithm gives up making $(j, i)$ valid. In particular,
no additional big jobs will be repelled.

Finally, we want to highlight the following counter-intuitive (but intentional) aspect of the algorithm.
It might happen that some job of size greater
than $W_0$ is moved to $i$,
only to be removed again in a later iteration,
when $W_0$ has increased.
\paragraph*{Critical move selection}
Suppose we are given some layers $L_{\le\ell+1}$ and big job moves $(j, i) \in L^B_{\mathrm{new}}$
where $j$ is repelled
by $\sigma(j)$ w.r.t. $L_{\le\ell+1}$, but not by $i$. Which of these moves should be
critical? Recall that for critical moves $(j, i)$ the target machine $i$ always repels
all jobs.
As in the exponential time algorithm, we later need to amortize these moves with small job moves.
Hence, we should select critical moves in a way that they produce many small job moves.

\begin{figure}
\centering
\begin{tikzpicture}
  \draw[thick] (-2, 0) rectangle (-1.75, 1);

  \draw[very thick] (2, 1.5) -- (2, -0.1) -- (2.5, -0.1) -- (2.5, 1.5);
  \draw[thick] (2.1, 0) rectangle (2.4, 0.1);
  \draw[thick] (2.1, 0.1) rectangle (2.4, 0.2);
  \draw[thick] (2.1, 0.2) rectangle (2.4, 0.3);
  \draw[thick] (2.1, 0.3) rectangle (2.4, 0.4);
  \draw[thick] (2.1, 0.4) rectangle (2.4, 0.5);
  \draw[thick] (2.1, 0.5) rectangle (2.4, 0.6);
  \draw[thick] (2.1, 0.6) rectangle (2.4, 0.7);
  \draw[thick] (2.1, 0.7) rectangle (2.4, 0.8);
  \draw[->, thick] (-1.9, 0.5) -- (1.9, 0.5) node[above, pos=0.5] {can move here};
  \draw[->, thick] (2.2, 0.5) -- (3.9, -0.4);

  \node at (-1.9, -0.4) {$\vdots$};
  \draw[->, thick] (2.5, -0.5) -- (3.9, -0.5);
  \node at (2.2, -0.4) {$\vdots$};

  \draw[thick] (-2, -2.5) rectangle (-1.75, -1.5);

  \draw[very thick] (2, -1) -- (2, -2.6) -- (2.5, -2.6) -- (2.5, -1);
  \draw[thick] (2.1, -2.5 + 0) rectangle (2.4, -2.5 + 0.1);
  \draw[thick] (2.1, -2.5 + 0.1) rectangle (2.4, -2.5 + 0.2);
  \draw[thick] (2.1, -2.5 + 0.2) rectangle (2.4, -2.5 + 0.3);
  \draw[thick] (2.1, -2.5 + 0.3) rectangle (2.4, -2.5 + 0.4);
  \draw[thick] (2.1, -2.5 + 0.4) rectangle (2.4, -2.5 + 0.5);
  \draw[thick] (2.1, -2.5 + 0.5) rectangle (2.4, -2.5 + 0.6);
  \draw[thick] (2.1, -2.5 + 0.6) rectangle (2.4, -2.5 + 0.7);
  \draw[thick] (2.1, -2.5 + 0.7) rectangle (2.4, -2.5 + 0.8);
  \draw[->, thick] (-1.9, -2) -- (1.9, -2);
  \draw[->, thick] (2.2, -2) -- (3.9, -0.6);

  \draw[very thick] (4, 0.5) -- (4, -1.1) -- (4.5, -1.1) -- (4.5, 0.5);

  \draw[thick, dashed] (-1.9, -0.6) ellipse (0.75cm and 2.2cm);
  \node at (-4.5, 0) {Big jobs repelled};
  \node at (-4.5, -0.5) {by their machines};
\end{tikzpicture}
\caption{Bottleneck for small jobs}
\label{sm-bottleneck}
\end{figure}
In the following let $\overline \machines$ denote the set of machines that repel all jobs
w.r.t. $L_{\le \ell+1}$.
As a prime example of a situation we want to avoid, consider the following:
There are a lot of critical moves $(j, i)$ where $p_j=1$, but
on $i$ there is a load of small jobs with volume
slightly above $(11/6 + 2\epsilon)\tau - p_j = (5/6 + 2\epsilon)\tau \ll \tau$ .
Moreover, these small jobs $j_S$ cannot go anywhere (meaning later layers will not have moves
for them), because
all their potential machines are in $\overline\machines$, i.e.,
$\Gamma(j_S)\setminus\{i\} \subseteq \overline \machines$.
Hence, there will not be any machines to amortize this
low load.
We should consider small jobs like this as blocked volume and when there is too much
blocked volume on $i$, a move $(j, i)$ should not be critical.
However, it is not enough to consider small jobs that have nowhere to go. 
It might also be that a lot of small jobs have only very few machines to go to.
In the example above, imagine that all the small jobs share only one machine $i'\notin\overline\machines$ to which they could go (see Fig.~\ref{sm-bottleneck}).
Then the average load is still very low. This is also something we want to avoid.
So how do we avoid these situations?
We will make sure that for 
non-valid critical moves $(j, i)$ where $j$ is big, $i$ has
one or more private machine $i'\in\overline\machines$,
which are reachable by some of its small jobs.
We select the critical edges sequentially (see Alg.~\ref{alg:critical}).
For the already added critical moves $(j, i)$, we consider all machines reachable by a small job on $i$
can go to as blocked as well, i.e., we add them to $\overline\machines$. 
We add $(j, i)$ to the critical moves only when the blocked small jobs (as described above)
and the medium jobs on $i$ have a volume that allows $j$ to be added to $i$, if there
were no other jobs.

\begin{algorithm}
\caption{Selection of critical moves}
\label{alg:critical}
\begin{algorithmic}
\STATE{$C \gets \emptyset$}
\STATE{let $\overline\machines$ be the set of the machines that repel all jobs w.r.t. $L_{\le\ell+1}$}
\FOR{$(j, i)\in L^B_{\mathrm{new}}$ ordered by $\pi$}
  \STATE{$\mathrm{Small} \gets p(\{j'\in\sigma^{-1}(i) : p_{j'} \le 1/2 \text{ and } \Gamma(j')\setminus\{i\} \subseteq \overline\machines\}$}
  \STATE{$\mathrm{Med} \gets p(\{j'\in\sigma^{-1}(i) : 1/2 < p_{j'} \le 5/6 \})$}
  \IF{$i\notin \overline\machines$ and
      $\mathrm{Small} + \mathrm{Med} + p_{j} \le 11/6 + 3\epsilon$}
   \STATE{$C\gets C\cup\{(j, i)\}$}
   \STATE{$\overline\machines\gets \overline\machines\cup \{i\}$}
   \FOR{$j'\in \sigma^{-1}(i)$ small}
     \STATE{$\overline\machines\gets \overline\machines\cup \Gamma(j')$}
   \ENDFOR
  \ENDIF
\ENDFOR
\RETURN $C$
\end{algorithmic}
\end{algorithm}

\subsection{Analysis}
\begin{lemma}
  If the configuration LP is feasible for $\tau$ and there remains a bad machine,
  then within the first $\ell \le 4\lceil\log_{1+\epsilon}(4|\machines|)\rceil$ layers
  there will be a valid leap or move.
\end{lemma}
\begin{proof}
Suppose toward contradiction, there are bad machines, no move in $L_{\le\ell}$ is valid, and $\ell = 4\lceil\log_{1+\epsilon}(4|\machines|)\rceil$.
We will construct values $(z_j)_{j\in\jobs}$, $(y_i)_{i\in\machines}$ with the properties as
in Lemma~\ref{lemma:condition-ra} and thereby show that the configuration LP is infeasible.
Throughout the proof the allocation $\sigma$ refers to the allocation
in the iteration where no move or leap is found.

First, we define values $z^{(k)}_j, y^{(k)}_i$ for all prefixes
of the layers ending in a leap layer,
i.e., for each $L_{\le 4k+1}$ with $0\le k < \ell/4$.
Furthermore, for technical reasons we define 
the values $z^{(-1)}_j, y^{(-1)}_i$ as well as $y^{(\ell/4)}_i$.
Then $z_j$, $y_i$ will be set as a positive linear combination of these values.

Let $\tilde\jobs(L_{\le 4k+1})$ denote all jobs $j$
that are repelled by $\sigma(j)$ w.r.t. $L_{\le 4k+1}$.
For every $0 \le k < \ell/4$ and $j\in\jobs$ let 
\begin{equation*}
  z^{(k)}_j = \begin{cases}
    \min\left\{\frac{p_j}{\tau}, \frac 5 6\right\} &\text{ if $j\in \tilde\jobs(L_{\le 4k+1})$}, \\
    0 &\text{ otherwise}.
  \end{cases}
\end{equation*}
Moreover, let $y^{(k)}_i := 1+\epsilon$,
if $i$ repels all jobs w.r.t. $L_{\le 4k+1}$ and
$y^{(k)}_i := z^{(k)}(\sigma^{-1}(i))$, otherwise.
Finally, define the corner cases $y^{(-1)}_i = 0$, $z^{(-1)}_j = 0$, and
$y^{(\ell/4)}_i := 1+\epsilon$ for all $i, j$.

Notice that $z^{(-1)}_j \le z^{(0)}_j \le \cdots \le z^{(\ell/4)}_j$ for all $j$ (and the same holds
for all $y^{(k)}_i$).
We set
\begin{align*}
  z^{(\le k)}_j &= \sum_{k'=-1}^k \frac{1}{(1+\epsilon)^{k'}}\cdot z^{(k')}_j , \\
  y^{(\le k)}_i &= \sum_{k'=-1}^{k} \frac{1}{(1+\epsilon)^{-k'}}\cdot y^{(k')}_i .
\end{align*}
The coefficients decrease exponentially with the layer number.
As we will see, this makes to the last values negligibly small (as in
the first algorithm).
Finally, set $z_j = z^{(\le\ell/4-1)}_j$ and $y_i = y^{(\le\ell/4)}_i$.
\begin{claim}\label{claim:constructive-feasibility}
Let $-1 \le k < \ell/4$, $i\in\machines$ and
$C\in\mathcal C(i, \tau)$.
Then
\begin{equation*}
  z^{(\le k)}(C) \le y^{(\le k+1)}_i .
\end{equation*}
\end{claim}
In particular, this implies $z(C) = z^{(\le\ell/4-1)} \le y^{(\le\ell/4)}_i = y_i$ for all $i, C$.
\begin{claim}\label{claim:constructive-unbounded}
\begin{equation*}
  \sum_{j\in\jobs} z_j > \sum_{i\in\machines} y_i .
\end{equation*}
\end{claim}
Together the claims imply $\tau < \OPT^*$, a contradiction.
\end{proof}
Before we prove the claims, we will state the following auxiliary lemmata.
\begin{lemma}\label{lemma:qp-aux1}
In an iteration where no valid move or leap is found
consider the set $L^B_{\mathrm{new}}$ selected in the algorithm after a leap layer $L_{\ell+1}$
is created
and let $(j_B, i)\in L^B_{\mathrm{new}}$.
Then
\begin{equation*}
  p\left(\left\{j\in \sigma^{-1}(i) : p_j \le \frac 5 6 \right\}\right) + p_{j_B} > \left(\frac{11}{6} + 2\epsilon\right)\tau .
\end{equation*}
\end{lemma}
\begin{proof}
Suppose toward contradiction that this does not hold.
Then $(j_B, i)$ is in the leap graph $G(\sigma)$.
It is also in $G(L_{\le\ell}, \sigma)$, since $i$ does not repel $j_B$ w.r.t. $L_{\le\ell+1}$.
Otherwise, $(j_B, i)$ would not be in $L^B_{\mathrm{new}}$.
Obviously $i$ is reachable by $j_B$ in $G(L_{\le\ell}, \sigma)$.
We argue that $i$ is reachable by some big job repelled by its current machine w.r.t. $L_{\le\ell}$.
This implies that $i$ repels $j_B$ w.r.t. $L_{\le\ell+1}$ by definition of repelled edges
for a leap layer. This is a contradiction, since $(j_B, i)$ could not be in $L^B_{\mathrm{new}}$ then.
We know that $\sigma(j_B)$ repels $j_B$ w.r.t. $L_{\le \ell+1}$.
If it repels $j_B$ already w.r.t. $L_{\le \ell}$, this follows trivially.
Otherwise, $j_B$ is repelled by $i$ because of the (leap) rule in the definition of repelled edges
for $L_{\ell+1}$. This can only be when $i\in L_{\ell+1}$, which means it is reachable by some big job
repelled by its machine w.r.t. $L_{\le \ell}$.
\end{proof}
\begin{lemma}\label{lemma:W0-bound}
In an iteration where no valid move or leap is found
consider the set $L^B_{\mathrm{new}}$ selected in the algorithm after a leap layer $L_{\ell+1}$
is created
and let $(j_B, i)\in L^B_{\mathrm{new}}$.
Then
\begin{equation*}
  p\left(\left\{j\in \sigma^{-1}(i) : \frac 1 2 < p_j \le \frac 5 6 \right\}\right) + p(S_i(L_{\le \ell+3},\sigma)) + p_{j_B} > \left(\frac{11}{6} + 2\epsilon\right)\tau ,
\end{equation*}
\end{lemma}
This lemma implies that the threshold $W_0$ chosen in the definition of repelled edges
always exists and $W_0\le 5/6$.
\begin{proof}
When $(j_B, i)$ is a critical moves in $L_{\ell+2}$ or when $(j_S, i)\in L_{\ell+3}$ for some small
job $j_S$, this is follows easily from the previous lemma,
since $S_i(L_{\le \ell+3},\sigma)$ contains all small jobs in $\sigma^{-1}(i)$.
Each other move $(j_B, i)$ would have been selected as a critical
move, if this inequality did not hold: Consider the set $\overline\machines$ in the selection
of critical moves at the time $(j_B, i)$ is considered. The algorithm adds $(j_B, i)$
to the critical moves, if $i\notin\overline\machines$ and
\begin{equation}
  p\left(\left\{j\in \sigma^{-1}(i) : \frac 1 2 < p_j \le \frac 5 6 \right\}\right)
  + p(S'_i(\overline\machines, \sigma)) + p_{j_B} \le \left(\frac{11}{6} + 2\epsilon\right)\tau ,\label{criterion-critical}
\end{equation}
where $S'_i(\overline\machines, \sigma)$ is the set of small jobs $j_S\in\sigma^{-1}(i)$ with
$\Gamma(j_S)\setminus\{i\}\subseteq \overline\machines$.
Recall that all machines in $\overline\machines$ either repel all jobs w.r.t. $L_{\le\ell+1}$ or
they are reachable by a small job on a machine that is target of a critical move. The latter kind
must repel all jobs w.r.t. $L_{\le\ell+3}$, because it is target of a small job move in $L_{\ell+3}$.
Thus, $S'_i(\overline\machines, \sigma) \subseteq S_i(L_{\le \ell+3},\sigma)$ and (\ref{criterion-critical})
is satisfied. Furthermore, $i\notin\overline\machines$, since $i$ does not repel
all jobs w.r.t. $L_{\le\ell+1}$ and we assumed
that there is no small job $j_S$ with $(j_S, i)\in L_{\ell+3}$.
Thus, $(j_B, i)$ would have been selected as a critical move.
\end{proof}
\begin{proof}[Proof of Claim~\ref{claim:constructive-feasibility}]
We argue inductively.
The basis of the induction is trivial, since $z^{(\le -1)}(C) = 0 \le y^{(\le 0)}_i$.
Suppose that $k\ge 0$ and for all $k' < k$,
\begin{equation*}
  z^{(\le k')}(C) \le y^{(\le k' + 1)}_i
\end{equation*}
If $y^{(k+1)}_i \ge 1+\epsilon$ then immediately
\begin{multline*}
  z^{(\le k)}(C) = z^{(\le k-1)}(C) + (1+\epsilon)^{-k} z^{(k)}(C) \le y^{(\le k)}_i + (1+\epsilon)^{-k} \frac{p(C)}{\tau} \\
  \le y^{(\le k)}_i + (1+\epsilon)^{-(k+1)} y^{(k+1)}_i = y^{(\le k+1)}_i .
\end{multline*}
We can therefore assume w.l.o.g. that $y^{(k+1)}_i = z^{(k+1)}(\sigma^{-1}(i))$.
Thus, $k<\ell/4$ and $i$ does not repel all jobs w.r.t. $L_{\le 4(k+1)+1}$.
Since by definition of repelled jobs, a machine that repels any small
job from another machine always repels all jobs, we
know that $i$ does not repel small jobs that are on other machines
w.r.t. $L_{\le 4(k+1)+1}$. Hence, for all small jobs $j_S\in C\setminus\sigma^{-1}(i)$ it holds that $z^{(k)}_{j_S} = 0$: If this was not true,
$\sigma(j_S)$ would repel $j_S$ w.r.t. $L_{4k+1}$,
in which case $(j_S, i)$ would have been added to
$L_{4k+3}$ and $i$ would repel all jobs, which is not true.

Consider the cases of big jobs in $C$.
If there is none, then obviously every jobs in $C\setminus\sigma^{-1}(i)$ is small.
Let $k' \le k$.
Then for all $j\in C\setminus\sigma^{-1}(i)$ it holds that $z^{(k')}_j \le z^{(k)}_j = 0$.
Consequently, $z^{(k')}(C) \le z^{(k')}(\sigma^{-1}(i)) = y^{(k')}_i$.
Hence,
\begin{equation*}
  z^{(\le k)}(C) \le z^{(\le k)}(\sigma^{-1}(i)) = y^{(\le k)}_i \le y^{(\le k+1)})_i .
\end{equation*}
Clearly, there can be at most one big job $j_B\in C$,
since such a job has $p_{j_B} > 1/2 \cdot \tau$ and $C$ cannot have a volume greater than $\tau$.
If $z^{(k)}_{j_B} = 0$ or $j_B\in \sigma^{-1}(i)$, the argument above still works.

We recap: The crucial case is when $y^{(k+1)}_i = z^{(k+1)}(\sigma^{-1}(i))$,
there is exactly one big job $j_B \in C\setminus\sigma^{-1}(i)$,
and $z^{(k)}_{j_B} = \min\{p_{j_B}/\tau, 5/6\}$.
Let $k'\le k$ be minimal with $z^{(k')}_{j_B} = \min\{p_{j_B}/\tau, 5/6\}$.
In particular, $z^{(-1)}_{j_B} = z^{(0)}_{j_B} = \cdots = z^{(k'-1)}_{j_B} = 0$.

\begin{description}
\item[Case 1: $i$ repels $j_B$ w.r.t. $L_{\le 4k'+1}$.]
This can either be because of a leap layer or a move layer in $L_{\le 4k'+1}$.
In the former case, there has to be a huge job in $j_H\in\sigma^{-1}(i)$ which $i$
repels w.r.t. $L_{\le 4k'+1}$. Otherwise, there would be a valid path.
Thus, for all $k'' \ge k'$ it holds that $z^{(k'')}_{j_H} = 5/6 \ge z^{(k'')}_{j_B}$ and
\begin{multline*}
  z^{(k'')}(C) = z^{(k'')}_{j_B} + z^{(k'')}(C\setminus\{j_B\})
  \le z^{(k'')}_{j_H} + z^{(k'')}(\sigma^{-1}(i)\setminus\{j_H\}) \\
  \le z^{(k'')}(\sigma^{-1}(i)) \le y^{(k'')}_i .
\end{multline*}
Furthermore, for all $k'' < k'$,
\begin{equation*}
  z^{(k'')}(C) = z^{(k'')}(C\setminus\{j_B\})
  \le z^{(k'')}(\sigma^{-1}(i)) \le y^{(k'')}_i .
\end{equation*}
Hence,
$z^{(\le k)}(C) \le y^{(\le k)}_i \le y^{(\le k+1)}_i$.

Now consider the case in which there is some move $(j_{k'}, i)$
which causes $i$ to repel $j_B$ is w.r.t. $L_{\le 4 k'+1}$.
The move $(j_{k'}, i)$ must be in a non-critical layer $L_{4k''+4}$, where $k'' < k'$,
since $i$ does not repel all jobs w.r.t. $L_{\le 4 k'+1}$.
Let $W_0$ as in the definition of repelled jobs in consequence of $(j_{k'}, i)$
and let
\begin{equation*}
  R = \{j\in\sigma^{-1}(i) : 1/2 < p_j \le W_0\}\cup S_i(L_{\le 4(k''-1)+3},\sigma) ,
\end{equation*}
The edges repelled by $i$ because of $(j_{k'}, i)$ are exactly $S_i(L_{\le 4k''+3},\sigma)$ and
all those $j$ with $1/2 < p_j \le W_0$. Hence, $p_{j_B} \le W_0$.
Recall that $W_0$ is chosen minimal with $p(R) + p_{j_{k'}} > (11/6 + 2 \epsilon)\tau$.
There must be a job $j'_B\in\sigma^{-1}(i)$ with $p_{j'_B} = W_0$, since otherwise
 $W_0$ would not be minimal.
Thus, for all $k'''$ it holds that $z^{(k''')}_{j'_B} \ge z^{(k''')}_{j_B}$ and
\begin{equation*}
  z^{(k''')}(C) = z^{(k''')}_{j_B} + z^{(k''')}(C \setminus \{j_B\})
  \le z^{(k''')}_{j'_B} + z^{(k''')}(\sigma^{-1}(i) \setminus \{j'_B\})
  \le y^{(k''')}_i .
\end{equation*}
It follows conveniently that $z^{(\le k)}(C) \le y^{(\le k)}_i \le y^{(\le k+1)}_i$.

\item[Case 2: $i$ does not repel $j_B$ w.r.t. $L_{\le 4k'+1}$]
Since $z^{(k')}_{j_B} > 0$, $j_B$ is repelled by $\sigma(j_B)$
w.r.t. $L_{\le 4k' + 1}$. Machine $i$ does not repel all jobs w.r.t. $L_{\le 4k'+1}$, which
implies there is no move with target $i$ in $L_{4k' + 2}$ or $L_{4k' + 3}$.
Hence, $(j_B, i)$ must be a move in layer $L_{4k' + 4}$.
Let
\begin{equation*}
  R = \{j\in\sigma^{-1}(i) : 1/2 < p_j \le W_0\}\cup S_i(L_{\le 4k'+1}, \sigma) ,
\end{equation*}
where $W_0$ is as in the definition of repelled jobs in consequence of $(j_B, i)$. Then
\begin{equation*}
  p(R) + p_{j_B} > \left(\frac{11}{6} + 2\epsilon\right) \tau \ge p(C) + \left(\frac 5 6 + 2\epsilon\right) \tau .
\end{equation*}
Furthermore, all jobs in $R$ are repelled by $i$ w.r.t. $L_{\le 4k' + 4}$ and therefore in $\tilde\jobs(L_{\le 4(k' + 1) + 1})$.
Since for all $j'\in R$ it holds that $p_{j'} \le W_0 \le 5/6$ (see Lemma~\ref{lemma:W0-bound}),
it follows that $z^{(k'' + 1)}_{j'} = p_{j'} / \tau$ for all $k'' \ge k'$.
Thus,
\begin{align*}
  z^{(k'')}(C) &= z^{(k'')}_{j_B} + z^{(k'')}(C\setminus\{j_B\}) \\
  &\le z^{(k'')}_{j_B} + (p(C) - p_{j_B}) / \tau \\
  &< z^{(k'')}_{j_B} + (p(R) - (5/6 + 2\epsilon) \tau) / \tau \\
  &= z^{(k'')}_{j_B} + p(R)/\tau - 5/6 - 2\epsilon \\
  &\le (1 - \epsilon) p(R) / \tau \\
  &\le (1-\epsilon) z^{(k'' + 1)}(\sigma^{-1}(i))
  \le \frac{z^{(k'' + 1)}(\sigma^{-1}(i))}{1+\epsilon} .
\end{align*}
Here we use that $i$ is a good machine and therefore
$p(R)\le p(\sigma^{-1}(i)) \le (11/6 + 2\epsilon)\tau < 2\tau$.
We conclude,
\begin{align*}
  z^{(\le k)}(C) &= z^{(\le k' - 1)}(C)
  + \sum_{k'' = k'}^k (1 + \epsilon)^{-k''} z^{(k'')}(C) \\
  &\le y^{(\le k')}_i
  + \sum_{k'' = k'}^k (1 + \epsilon)^{-k''} \frac{z^{(k''+1)}(\sigma^{-1}(i))}{1+\epsilon}  \\
  &\le y^{(\le k')}_i
  + \sum_{k'' = k'}^k (1 + \epsilon)^{-(k'' + 1)} y^{(k'' + 1)}_i
  = y^{(\le k + 1)}_i . \qedhere
\end{align*}
\end{description}
\end{proof}
\begin{proof}[Proof of Claim~\ref{claim:constructive-unbounded}]
Let $i$ be a bad machine.
Then $i$ repels all jobs (in particular those in $\sigma^{-1}(i)$)
w.r.t. $L_{\le 0}$.
Hence, for every $0\le k < \ell/4$ and $j\in\sigma^{-1}(i)$,
$z^{(k)}_j = \min\{5/6, p_j/\tau\} \ge 5/6 \cdot p_j/\tau$
Thus,
\begin{equation*}
  z^{(k)}(\sigma^{-1}(i)) \ge \frac 5 6 p(\sigma^{-1}(i))/\tau > \frac 5 6 \left(\frac {11} 6 + 2\epsilon\right) > \frac{55}{36} + \epsilon > 1+\epsilon + \frac 1 2 .
\end{equation*}
This implies
\begin{multline*}
  y_i = \sum_{k=0}^{\ell/4} (1+\epsilon)^{-k} y^{(k)}_i = \sum_{k=0}^{\ell/4-1}[(1+\epsilon)^{-k} (1+\epsilon)] + (1+\epsilon)^{-(\ell/4-1)} \\
  < \sum_{k=0}^{\ell/4-1} [(1+\epsilon)^{-k} \cdot z^{(k)}(\sigma^{-1}(i))] + (1+\epsilon)^{-(\ell/4-1)} - \frac 1 2 \sum_{k=0}^{\ell/4-1} (1+\epsilon)^{-k} \\
  \le z(\sigma^{-1}(i)) - \frac 1 2 .
\end{multline*}
In the last inequation, we use the last two elements of the sum
$\sum_{k=0}^{\ell/4-1} (1+\epsilon)^{-k}$ to compensate for $(1+\epsilon)^{-(\ell/4-1)}$.
The inequation shows that $y_i$ is much smaller than $z(\sigma^{-1}(i))$.
If for all good machines $i$ and layers $k$ we had $y^{(k)}_i = z^{(k)}(\sigma^{-1}(i))$
(which is the case when $i$ does not repel all jobs w.r.t. $L_{\le k}$),
the proof would be easy: $\sum_{j\in\jobs} z^{(\le\ell/4-1)}_j = \sum_{i\in\machines}z^{(\le\ell/4-1)}(\sigma^{-1}(i))$
would be larger than $1/2 + \sum_{i\in\machines} y^{(\le\ell/4-1)}_i$.
The former is exactly $\sum_{j\in\jobs} z_j$ and the latter is
\begin{equation*}
1/2 + \sum_{i\in\machines} y_i - \sum_{i\in\machines} (1+\epsilon)^{-\ell/4} y^{(\ell/4)}_i < \sum_{i\in\machines} y_i .
\end{equation*}
Here we use that the decrease in the coefficient
makes $y^{(\ell/4)}_i$ neglectable, which we will explain in detail as we go
through the actual proof.

Of course, there can be machines that repel all jobs and
are set to $y^{(k)}_i = 1+\epsilon$.
We have to make sure that they do not have a negative effect.
Let $B_k$ be the machines $i$ with $(j_B, i)$ in the $k$-th critical layer for some $j_B$,
i.e., in $L_{4k+1}$.
Let $A_k$ be the machines $i$ with $(j_S, i)$ in the $k$-th small layer for some $j_S$,
i.e., in $L_{4k+3}$.

Let $k<\ell/4$ and $i\in B_k$. $i$ repels all jobs w.r.t. $L_{\le 4(k+1)+1}$. Thus,
$\sigma^{-1}(i)\subseteq \tilde\jobs(L_{\le 4(k+1)+1})$.
Let $(j_B, i)$ as above. This move is not valid. 
Either there is a job $j\in\sigma^{-1}(i)$ with $z^{(k+1)}_j = 5/6$ or $z^{(k+1)}_j = p_j / \tau$ for all $j\in\sigma^{-1}(i)$. Thus,
\begin{multline*}
  z^{(k+1)}(\sigma^{-1}(i)) \ge \min\{5/6,\ p(\sigma^{-1}(i)) / \tau\} \\
  \ge \min\{5/6,\ 11/6 + 2\epsilon - p_{j_B} / \tau\}
  \ge 5/6 \ge 1 + \epsilon - \epsilon - 1/6
  = y^{(k+1)}_i - \epsilon - 1/6 .
\end{multline*}
Next, let $i\in A_k$. Then there is a move $(j_S, i) \in L_{\le 4k+1}$ with $j_S$ small. Of course, this move is not valid either. In the following, we distinguish between the cases where $\sigma^{-1}(i)$ has no huge job
or one huge job.
\begin{multline*}
  z^{(k+1)}(\sigma^{-1}(i)) \ge \min\{p(\sigma^{-1}(i)) / \tau,\ (p(\sigma^{-1}(i)) - \tau) / \tau + 5/6\} \\
  \ge 8/6 + 2\epsilon - 1 + 5/6
  = \frac 7 6 + 2\epsilon \ge y^{(k+1)}_i + 1/6 + 2\epsilon .
\end{multline*}
The bounds above show that machines in $A_k$ have $z^{(k+1)}(\sigma^{-1}(i))$ above $y^{(k+1)}_i$
and machines in $B_k$ below.
In order to amortize the machines, we have to proof a bounded ratio between them:
We argue that for every $k < \ell/4$, $|A_{k}| \ge |B_{k}|$.
Notice that $B_{k} \dot\cup A_{k}$ are exactly the machines that are added to $\overline\machines$
in the selection of critical moves for $L_{4k+2}$.
Hence, it suffices to show that at most half of them are target of critical big job moves.
Consider a critical move $(j_B, i)$ for a big job $j_B$ that is added in the critical move selection.
By Lemma~\ref{lemma:qp-aux1}
\begin{equation*}
  p\left(\left\{j\in\sigma^{-1}(i) : p_j \le \frac 5 6 \right\}\right) + p_{j_B} > \left(\frac{11}{6} + 2\epsilon\right)\tau .
\end{equation*}
Because the move $(j_B, i)$ is selected as a critical move, it holds that
\begin{equation*}
  p\left(\left\{j\in\sigma^{-1}(i) : \frac 1 2 < p_j \le \frac 5 6 \right\}\right) + p(S'_i(\overline\machines, \sigma)) + p_{j_B} \le \left(\frac{11}{6} + 2\epsilon \right)\tau ,
\end{equation*}
where $S'_i(\overline\machines, \sigma)$ are the small jobs $j_S\in\sigma^{-1}(i)$
with $\Gamma(j_S)\setminus\{i\}\subseteq \overline\machines$ with $\overline\machines$ as
at the time before $(j_B, i)$ is selected.
Consequently, there is a small job $j_S\in\sigma^{-1}(i)\setminus S'_i(\overline\machines, \sigma)$.
It follows that there exists a machine $i'\in\Gamma(j_S)\setminus (\overline\machines\cup\{i\})$.
The algorithm adds $i$ and $i'$ to $\overline\machines$.
In other words, whenever the algorithm adds a machine $i$ to $B_k$, it adds at least one machine
$i'$ to $A_k$.
It follows that
\begin{align*}
  \sum_{j\in\jobs} z_j 
  &= \sum_{k=0}^{\ell/4-1}\sum_{i\in\machines}(1+\epsilon)^{-k} z^{(k)}(\sigma^{-1}(i)) \\
  &> \sum_{k=0}^{\ell/4-1} (1+\epsilon)^{-k} \bigg[ \left(\frac 1 6+\epsilon\right)\underbrace{(|A_{k}| - |B_k|)}_{\ge 0} + \sum_{i\in\machines} y^{(k)}_i \bigg] + \frac 1 2 \\
  &\ge \sum_{i\in\machines} y_i
  + \underbrace{\frac 1 2 - \sum_{i\in\machines} (1+\epsilon)^{-\ell/4} y^{(\ell/4)}_i}_{\ge 0}
  \ge \sum_{i\in\machines} y_i
\end{align*}
In the last inequality we use that
by choice of $\ell$, $(1+\epsilon)^{\ell/4} \le 4 |\machines|$, which implies
\begin{equation*}
  \sum_{i\in\machines} (1+\epsilon)^{-\ell/4} y^{(\ell/4)}_i
  \le 2 |\machines|(1+\epsilon)^{-\ell/4} \le \frac 1 2 . \qedhere
\end{equation*}
\end{proof}
\begin{lemma}
  The algorithm terminates in time $n^{O(1/\epsilon \log(n))}$, where $n = |\jobs| + |\machines|$.
\end{lemma}
\begin{proof}
  We are looking at the states of two consecutive iterations right before a move or leap
  is performed.
  Let $\sigma$ be the schedule in the former iteration and $\sigma'$ in the latter.
  Likewise, define layers $L_{\le\ell}$ and $L'_{\le\ell'}$
  right before they collapse.
  Let $\tilde\jobs(L_{\le k},\sigma)$ be the jobs $j$ repelled by $\sigma(j)$ w.r.t. $L_{\le k}$.
  Recall, $R(L_{\le k}, \sigma)$
  is the set of all $(j, i)\in \jobs\times\machines$ where $i$ repels $j$ w.r.t. $L_{\le k}$.

  We define a potential function $\Phi$ for each of the layers. Set
\begin{equation*}
  \Phi(L_{\le k},\sigma) = \begin{cases}
    |R(L_{\le k}, \sigma)| &\text{if $L_k$ is a leap layer}, \\
    (|L_k|, |\jobs| - |\tilde\jobs(L_{\le k},\sigma)|) &\text{if $L_k$ is a critical layer,} \\
    |\jobs| - |\tilde\jobs(L_{\le k},\sigma)| &\text{if $L_k$ is a small layer,} \\
    (|R(L_{\le k}, \sigma)|, |\jobs| - |\tilde\jobs(L_{\le k},\sigma)|) &\text{if $L_k$ is a non-critical layer}.
  \end{cases}
\end{equation*}

\begin{claim}\label{claim:lexicographic-decrease}
  The vector
  \begin{equation*}
    (g', |\jobs| - |\tilde\jobs(L'_{\le 0},\sigma')|, \Phi(L'_{\le 1},\sigma'), \dotsc, \Phi(L'_{\le\ell'},\sigma'), \infty)
  \end{equation*}
  is lexicographically bigger than
  \begin{equation*}
    (g, |\jobs| - |\tilde\jobs(L_{\le 0},\sigma)|, \Phi(L_{\le 1},\sigma), \dotsc, \Phi(L_{\le\ell},\sigma), \infty)
  \end{equation*}
\end{claim}
  Since the number of layers is at most $O(1/\epsilon \log(n))$
  and components can have only $O(n^3)$ different values,
  the number of vectors is
  is bounded by $n^{O(1/\epsilon \log(n))}$. Since for every move or leap it decreases lexicographically,
  the lemma follows easily from the claim.
\end{proof}
\begin{proof}[Proof of Claim~\ref{claim:lexicographic-decrease}]
  If the number of good machines increases, the claim follows immediately.
  If it does not, but a job is moved from a bad machine to a good one, then
  $\tilde\jobs'(L'_{\le 0},\sigma') \subsetneq \tilde\jobs(L_{\le 0},\sigma)$, i.e.,
the claim follows again.
  Hence, assume neither case is true.
  Let $1 \le k\le\min\{\ell-1,\ell'\}$ and:
\begin{enumerate}
  \item $L_{\le k-1} = L'_{\le k-1}$;
  \item $\tilde\jobs(L_{\le k-1}) = \tilde\jobs'(L'_{\le k-1})$ and $\sigma(j) = \sigma'(j)$ for all $j\in\tilde\jobs(L_{\le k-1})$;
  \item $R(L'_{\le k-1}, \sigma') = R(L_{\le k}, \sigma)$.
\end{enumerate}
  We will prove:
  $\Phi(L'_{\le k}) \ge \Phi(L_{\le k})$ and
  if equality holds, (1), (2) and (3) also hold for $k$.
  This implies the claim by induction:
  If $\ell' < \ell$, then the prefix of the first vector ending in $\Phi(L'_{\le \ell'},\sigma)$
  is lexicographically not smaller than the prefix of the second one ending in
  $\Phi(L_{\le \ell'},\sigma)$. Furthermore, the next component in the first vector is
  $\infty$, whereas it is something finite in the second.
  If $\ell' \ge \ell$, then we notice that (2) cannot hold for $k-1=\ell-1$. This is
  because some leap or move in $L_{\ell}$ was executed and therefore a job $j$ that
  is repelled by $\sigma(j)$ w.r.t. $L_{\le\ell-1}$ was moved.
\begin{description}
\item[Case 1: $L_k$ is a leap layer.]
  We argue that every machine $i$ reachable in the leap graph $G(L_{\le k-1}, \sigma)$ by a 
  job big job $j_0$ repelled by $\sigma(j_0)$ w.r.t. $L_{\le k-1}$ is also reachable by $j_0$ in the leap
  graph $G(L'_{\le k-1}, \sigma')$. Because of (2),
  this means that the set of reachable machines in $\sigma$
  by any such job, which is exactly $L_k$, is a subset of $L'_k$.
  It suffices to show that every edge reachable by $j_0$ in $G(L_{\le k-1}, \sigma)$ is
  also in the leap graph $G(L'_{\le k-1}, \sigma')$.
  Because of (3) it suffices to show that it is in $G(\sigma')$.
  An edge in the leap graph can be one of two kinds.
  It can be from a machine to a huge job, i.e., $(i, j_H)$,
  which exists because $\sigma(j_H) = i$.
  We argue that $j_H$ was not not moved, which means $\sigma'(j_H) = \sigma(j_H) = i$
  and therefore the edge is also in $G(\sigma')$.
  Suppose toward contradiction that a move $(j_H, i')$ was executed.
  Then there is no huge job in $\sigma^{-1}(i')$ and
  $p(\sigma^{-1}(i')) + p_{j_H} \le (11/6 + 2\epsilon) \cdot \tau$. Therefore $i'$
  would also be reachable by $j_0$ in $G(\sigma)$. Hence, the algorithm
  would execute a leap in $L_k$, which it did not, since $\ell > k$.
  Similarly, if $j_H$ was moved as part of a leap, then all machines in this leap
  would be reachable already in $L_k$ and there would have been a valid leap already.

  Now consider an edge of the form $(j_B, i)$. If $i$ had no huge job in $\sigma^{-1}(i)$, then
  again, there would have been a valid leap in $L_k$, which cannot be.
  The edge $(j_B, i)$ exists in the leap graph of $\sigma$
  because $i\in\Gamma(j_B)\setminus\{\sigma(j_B)\}$ and
\begin{equation*}
  p(\{j\in\sigma^{-1}(i) : p_{j} \le 5/6\}) + p_{j_B} \le (11/6 + 2\epsilon) \tau .
\end{equation*}
  It could only be removed, if $j_B$ was moved to $i$%
  ---this cannot be the case for the same reason as above---or
  some job $j'$ with $p_{j'}\le 5/6$ is moved to $i$. By definition of a valid non-huge move, however, this means that
\begin{multline*}
  (11/6 + 2\epsilon) \tau \ge p(\{j\in\sigma^{-1}(i) : p_{j} \le 5/6\}) + p_{j'} + 1 \\
  \ge p(\{j\in\underbrace{\sigma^{-1}(i)\cup \{p_{j'}\}}_{=\sigma^{\prime-1}(i)} : p_{j} \le 5/6\}) + p_{j_B} .
\end{multline*}
  Therefore, $(j_B, i)$ is also in $G(\sigma')$.
  We conclude, all reachable machines in $L_k$ are also in $L'_k$, i.e., $L'_k \supseteq L_k$.
  By the arguments above, every job adjacent to a machine
  in $L_k$ in $G(L_{\le k-1}, \sigma)$ is also adjacent
  to this machine in $L'_k$ in $G(L'_{\le k-1}, \sigma')$. Thus,
  $R(L'_{\le k},\sigma')\supseteq R(L_{\le k},\sigma)$.
  If equality does not hold, then $\Phi(L'_{\le k}, \sigma') > \Phi(L_{\le k}, \sigma)$.
  Otherwise, (3) holds for $k$. (1) must also hold for $k$, because $L'_k \supsetneq L_k$ was true,
  then the additional machine would repel at least one additional job (its huge job).
  Finally, (2) holds, because no huge job on a reachable machine was moved as elaborated above.
\item[Case 2: $L_k$ is a critical layer.]
  We show that every critical move in $L_k$ is also in $L'_k$.
  By induction hypothesis, we know that the moves $(j, i)$, $i\in\Gamma(j)$, where $j$ is a big job
  repelled by $\sigma(j) = \sigma'(j)$, but not by $i$,
   w.r.t. $L_{\le k-1}$ and w.r.t. $L'_{\le k-1}$ are the same.
  Therefore, the sets $L^B_{\mathrm{new}}$
  from which the critical moves are selected are the same in both cases.
  Recall that critical moves are added greedily in the order of $\pi'$ ($\pi$).
  In $\pi'$ the moves $L^B_{\mathrm{new}}$ are ordered in a way that first the moves from $L_{k}$ appear
  (in the order of $\pi$) and then all others. This is because in the main algorithm when $L_k$ was
  created, all $(j, i)\in L_k$ were moved to the front of $\pi$. We just have to understand that
  none of $L^B_{\mathrm{new}}\setminus L_k$ were moved to the front at a later time.
  This is because there is no way that a move, which is not selected as critical, can be selected in a
  later layer.

  Let $(j_1, i_1), \dotsc, (j_{r-1}, i_{r-1})$ be the first $r-1$ critical moves selected in $L_k$.
  Furthermore, let $\overline\machines_{r-1}$ and $\overline\machines_{r-1}'$
  be as in the algorithm before the $r$-th critical move was added.
  Note that before the first critical move was added, by (3) it holds that
  $\overline\machines'_{0} = \overline\machines_{0}$, since these are the machines
  that repel all jobs w.r.t. $L_{\le k-1}$.
  We assume for induction that $\overline\machines'_{r-1} \subseteq \overline\machines_{r-1}$ and
  that $(j_1, i_1), \dotsc, (j_{r-1}, i_{r-1})$ were also added to $L'_k$.
  Because no move or leap in $L_{\le k-1}$ was executed and $i_r$ repels all jobs w.r.t. $L_{\le k}$,
  we know that $\sigma^{\prime-1}(i_r)\subseteq \sigma^{-1}(i_r)$. In particular,
  every medium job
  in $\sigma^{\prime-1}(i_r)$ was already in $\sigma^{-1}(i_r)$. Moreover, every small job
  $j_S\in \sigma^{\prime-1}(i_r)$ with $\Gamma(j_S)\setminus\{i_r\}\subseteq\overline\machines'_{r-1} \subseteq \overline\machines_{r-1}$ was also in $\sigma^{-1}(i_r)$ . Hence, the condition for adding
  $(j_r, i_r)$ to $L'_k$ holds, since it did for $L_k$.
  Finally, $\overline\machines'_r$ is the union of $\overline\machines'_{r-1}$, $\{i\}$,
  and $\Gamma(j_S)$ for
  every small $j_S\in\sigma^{\prime-1}(i_r)$. This is a subset of $\overline\machines_{r-1}$, $\{i\}$,
  and $\Gamma(j_S)$ for every small $j_S\in\sigma^{-1}(i_r)\subseteq\sigma^{\prime-1}(i_r)$,
  which is $\overline\machines_r$.

  If $L'_k \supsetneq L_k$, nothing has to be shown, since $\Phi(L'_{\le k}) > \Phi(L_{\le k})$.
  Otherwise, $L'_k = L_k$ and therefore (3) follows for $k$ directly.
  If some job was moved away from a machine of a critical move, then again
  $\Phi(L'_{\le k}) > \Phi(L_{\le k})$. Otherwise, (2) follows for $k$.
\item[Case 3: $L_k$ is a small layer.]
  As in the previous case, we have that $L_k = L'_k$, i.e., (1) holds also for $k$.
  (3) also holds for $k$, since in the rules of a small layer, every target of a move
  repels every job. This is the same in $L'_{\le k}$ and $L_{\le k}$.
  If some job was moved away from
  a target machine of a move in $L_k$,
  then $\tilde\jobs(L'_{\le k}) \subsetneq \tilde\jobs(L_{\le k})$
  and therefore $\Phi(L'_{\le k}) > \Phi(L_{\le k})$. Otherwise, (2) follows for $k$ as well.
\item[Case 4: $L_k$ is a non-critical-layer.]
  By the arguments in Case 2 we know that the previous critical moves and the moves
  they are chosen from are the same
  and therefore also for the non-critical moves $L'_k = L_k$.
  Let $(j, i)\in L_k$.
  We argue that
  \begin{equation*}
    S_i(L_{\le k-1}, \sigma) = S_i(L'_{\le k-1}, \sigma') .
  \end{equation*}
  Let $j_S\in S_i(L_{\le k-1}, \sigma)$. Then there cannot be a move $(j_S, i')$ in
  some higher layer than $L_{k-1}$. This is because $j_S$ is repelled
  by all $i'\in\Gamma(j_S)\setminus\{\sigma(j_S)\}$ w.r.t.
  $L_{k-1}$. Hence, $\sigma'(j_S) = \sigma(j_S) = i$. With (3) it follows that
  $j_S\in S_i(L'_{\le k-1}, \sigma')$.
  Now let $j_S\in S_i(L'_{\le k-1}, \sigma')$. If $\sigma(j_S) = \sigma'(j_S) = i$, then
  as above with (3) it follows that $j_S\in S_i(L_{\le k-1}, \sigma)$.
  Now assume toward contradiction $\sigma(j_S) \neq i$.
  By (2),
  $j_S$ is not repelled by $\sigma(j_S)$ w.r.t. $L_{\le k-1}$;
  By (3) this means that $j_S$ is also not repelled by $\sigma(j_S)\neq i$ w.r.t.
  $L'_{\le k-1}$. Hence, $j_S\notin S_i(L'_{\le k-1}, \sigma')$, a contradiction.
  Let $W_0$ be the minimal $W\ge 0$ such that
\begin{equation*}
  p\left(\left\{j'\in\sigma^{-1}(i) : \frac 1 2 < p_{j'} \le W\right\}\right)
+ p(S_{i}(L_{\le k-1}, \sigma)) + p_{j}
  > \left(\frac{11}{6} + 2\epsilon\right) \tau .
\end{equation*}
  Since $i$ repels all jobs $j'$ with $1/2 < p_{j'} \le W$ w.r.t. $L_{\le k}$, we get
\begin{equation*}
  \left\{j'\in\sigma^{'-1}(i) : \frac 1 2 < p_{j'} \le W\right\}
  \subseteq \left\{j'\in\sigma^{-1}(i) : \frac 1 2 < p_{j'} \le W\right\} .
\end{equation*}
  This implies that $W'_0$, the minimal $W\ge 0$ with
\begin{equation*}
  p\left(\left\{j'\in\sigma^{\prime-1}(i) : \frac 1 2 < p_{j'} \le W\right\}\right)
+ p(S_{i}(L'_{\le k-1}, \sigma')) + p_{j}
  > \left(\frac{11}{6} + 3\epsilon\right) \tau .
\end{equation*}
  is at least as big as $W_0$, i.e., $W'_0\ge W_0$. This means all jobs repelled by $i$
  w.r.t. $L_{\le k}$ are also repelled w.r.t. $L'_{\le k}$, which implies
  $R(L'_{\le k}, \sigma') \supseteq R(L_{\le k}, \sigma)$. If equality does not hold,
  then $\Phi(L'_{\le k}, \sigma') > \Phi(L_{\le k}, \sigma)$.
  Otherwise (3) is fulfilled for $k$.
  If one of the jobs repelled by $i$ is moved,
  then $\tilde\jobs(L'_{\le k-1}) \subsetneq \tilde\jobs(L_{\le k-1})$. Otherwise,
  equality holds and (2) follows for $k$.
\end{description}
\end{proof}

\begin{theorem}
  We can find a $(11/6+\epsilon)$-approximate
  solution for \textsc{Restricted Assignment} in time $n^{O(1/\epsilon \log(n))}$
  for every $\epsilon > 0$, where $n = |\mathcal J| + |\mathcal M|$.
\end{theorem}

\bibliographystyle{plain}
\bibliography{main-plain}
\end{document}